\documentclass[sigconf, nonacm]{acmart}

\usepackage[]{hyperref}

\usepackage{tikz}
\usepackage{amsmath}

\usepackage{graphicx}

\usepackage{subcaption}

\usepackage{multirow}
\usepackage{rotating}
\usepackage{array}
\usepackage{balance}
\usepackage{enumitem}
\usepackage{xcolor}
\usepackage[ruled,linesnumbered,nofillcomment]{algorithm2e}
\usepackage{ulem}

\usepackage[subtle]{savetrees}
\usepackage{verbatim}
\usepackage{soul}
\usepackage{listings}
\usepackage{caption}
\usepackage{fontawesome}
\usepackage[dvipsnames]{xcolor}
\usepackage{afterpage}
\newlength{\oldtextfloatsep}
\definecolor{darkcerulean}{rgb}{0.03, 0.27, 0.49}
\definecolor{codegreen}{rgb}{0,0.6,0}
\definecolor{codegray}{rgb}{0.5,0.5,0.5}
\definecolor{codepurple}{rgb}{0.58,0,0.82}
\definecolor{backcolour}{rgb}{0.95,0.95,0.92}
\definecolor{ballblue}{rgb}{0.13, 0.67, 0.8}
\definecolor{blue}{rgb}{0.0, 0.23, 0.84}
\definecolor{cobalt}{rgb}{0.0, 0.28, 0.67}
\definecolor{coolblack}{rgb}{0.0, 0.18, 0.39}
\definecolor{darkcerulean}{rgb}{0.03, 0.27, 0.49}
\definecolor{readgreen}{rgb}{0, 0.398, 0.199}
\definecolor{writepurple}{rgb}{0.906, 0.5195, 0.906}
\definecolor{compactpurple}{rgb}{0.48, 0, 0.5078}
\definecolor{lightgreen}{rgb}{0.832, 0.90625, 0.828125}
\definecolor{circledarkblue}{HTML}{00356C}
\definecolor{circledarkgreen}{HTML}{00B900}
\definecolor{darkblue}{HTML}{003A77}
\definecolor{darkgreen}{HTML}{008A00}
\definecolor{darkyellow}{HTML}{ef875d}
\definecolor{darkred}{HTML}{cc261b}

\usepackage{tikz}
\let\oldnl\nl
\newcommand{\nonl}{\renewcommand{\nl}{\let\nl\oldnl}}

\newcommand*\circledg[1]{\protect \tikz[baseline=(char.base)]{
            \protect \node[shape=circle,draw=circledarkgreen,inner sep=0.5pt,fill=circledarkgreen,text=white, scale=0.9] (char) {#1};}}
\newcommand*\circledb[1]{\protect \tikz[baseline=(char.base)]{
            \protect \node[shape=circle,draw=circledarkblue,inner sep=0.5pt,fill=circledarkblue,text=white, scale=0.9] (char) {#1};}}

\newcommand{\specialcell}[2][c]{%
  \begin{tabular}[#1]{@{}c@{}}#2\end{tabular}}

\newcommand{\rcross}[0]{\textcolor{darkred}{\faTimesCircle}}
\newcommand{\gcheck}[0]{\textcolor{darkgreen}{\faCheckCircle}}

\newcommand{\insightbox}[2]{%
    \noindent
    \begin{minipage}{\columnwidth}
    {\setlength{\fboxsep}{2pt}%
    \fbox{\parbox{\dimexpr\columnwidth-2\fboxsep-2\fboxrule\relax}{\normalsize #2}}}%
    \end{minipage}%
}

\newcommand{\pname}[1]{RAGPerf{#1}}

\newcommand\vldbdoi{XX.XX/XXX.XX}

\newcommand\vldbavailabilityurl{}
\newcolumntype{N}{@{}m{0pt}@{}}

\begin{document}

\settopmatter{printfolios=true}

\title{\pname{}: An End-to-End Benchmarking Framework for Retrieval-Augmented Generation Systems
}

\author{Shaobo Li}
\authornote{Co-primary authors.}
\email{shaobol2@illinois.edu}
\affiliation{%
  \institution{University of Illinois}
  \streetaddress{}
  \city{Urbana}
  \state{IL}
  \country{USA}
  \postcode{}
}

\author{Yirui Zhou}
\authornotemark[1]
\email{yiruiz2@illinois.edu}
\affiliation{%
  \institution{University of Illinois}
  \streetaddress{}
  \city{Urbana}
  \state{IL}
  \country{USA}
  \postcode{}
}

\author{Yuan Xu}
\email{yuanxu4@illinois.edu}
\affiliation{%
  \institution{University of Illinois}
  \streetaddress{}
  \city{Urbana}
  \state{IL}
  \country{USA}
  \postcode{}
}

\author{Kevin Chen}
\email{kpchen2@illinois.edu}
\affiliation{%
  \institution{University of Illinois}
  \streetaddress{}
  \city{Urbana}
  \state{IL}
  \country{USA}
  \postcode{}
}


\author{Daniel Waddington}
\email{Daniel.Waddington@ibm.com}
\affiliation{%
  \institution{IBM Research}
  \streetaddress{}
  \city{}
  \state{}
  \country{USA}
  \postcode{}
}

\author{Swaminathan Sundararaman}
\email{swami@ibm.com}
\affiliation{%
  \institution{IBM Research}
  \streetaddress{}
  \city{}
  \state{}
  \country{USA}
  \postcode{}
}

\author{Hubertus Franke}
\email{frankeh@us.ibm.com}
\affiliation{%
  \institution{IBM Research}
  \streetaddress{}
  \city{}
  \state{}
  \country{USA}
  \postcode{}
}

\author{Jian Huang}
\email{jianh@illinois.edu}
\affiliation{%
  \institution{University of Illinois}
  \city{Urbana}
    \state{IL}
  \country{USA}
}


\begin{abstract}

We present the design and implementation of a 
RAG-based AI system benchmarking (\pname{})
framework for characterizing the system behaviors of RAG pipelines. To facilitate detailed profiling and fine-grained performance analysis, \pname{} decouples the RAG workflow into several modular components — embedding, indexing, retrieval, reranking, and generation. \pname{} offers the flexibility for users to configure the core parameters of each component and examine their impact on the end-to-end query performance and quality.  
\pname{} has a workload generator to model real-world scenarios by supporting diverse datasets (e.g., text, pdf, code, and audio), different retrieval and update ratios, and query distributions. \pname{} also supports different embedding models, major vector databases such as LanceDB, Milvus, Qdrant, Chroma, and Elasticsearch, as well as different LLMs for content generation. It automates the collection of performance metrics (i.e., end-to-end query throughput, host/GPU memory footprint, and CPU/GPU utilization) and accuracy metrics (i.e., context recall, query accuracy, and factual consistency). We demonstrate the capabilities of \pname{} through a comprehensive set of experiments and open source its codebase at GitHub~\footnote{The codebase of \pname{} is available at https://github.com/platformxlab/RAGPerf}. Our evaluation shows that \pname{} incurs negligible performance overhead.

\end{abstract}

\maketitle


\ifdefempty{\vldbavailabilityurl}{}{
\vspace{.3cm}
\begingroup\small\noindent\raggedright\textbf{PVLDB Artifact Availability:}\\
The source code, data, and/or other artifacts have been made available at \url{\vldbavailabilityurl}.
\endgroup
}

\section{Introduction}
\label{sec:intro}

The advancement of large language models (LLMs) has revolutionized intelligent data processing, enabling a vast amount of AI applications. However, the capabilities of LLMs are still limited in scenarios where user queries require domain-specific knowledge or private data. To this end, retrieval-augmented generation (RAG)~\cite{lewis2020rag,huang2024survey,gao2023retrieval} has been developed to supplement LLMs with external knowledge bases. With RAG, we can incorporate up-to-date knowledge into the content generation process in LLMs, providing accurate responses for a variety of specific domains such as scientific research, legal discovery, and financial analysis.


As we deploy a RAG pipeline for a domain-specific application, it is essential to understand its system bottleneck and design trade-offs. However, studying the performance implications and query quality under realistic deployment scenarios is hard and time-consuming, due to the complexity of the computing stack across the entire RAG pipeline. 
Moreover, a RAG pipeline usually involves multiple components (i.e., embedding, indexing, retrieval, reranking, and generation), and the interaction among these components makes it difficult for developers to decide the optimal configurations. 

Therefore, to address above challenges, it is desirable for developers to have a reproducible, end-to-end benchmarking framework for RAG-based AI systems. 
Unfortunately, existing RAG benchmarks failed to achieve this goal (see a detailed comparison in Table~\ref{tab:benchmarks}). Most existing RAG benchmarks focused on semantic metrics for a specific application~\cite{friel2024ragbench, lin2022truthfulqa, niu2024ragtruth, zhao2023felm, pipitone2024legalbench, wang2025omnieval, kim2024autorag}, few prior studies examined the performance behaviors of the RAG pipeline. They cannot capture the runtime performance and potential resource contentions in a fully integrated RAG pipeline. And many of them lack the flexibility, and do not allow users to customize the RAG pipeline with different datasets, embedding models, and even vector databases. 

Inspired by benchmark frameworks like YCSB~\cite{ycsb} or OLTPbench~\cite{oltp} for cloud services, we argue that the RAG ecosystem requires a dedicated benchmarking framework to characterize the end-to-end RAG performance. In this paper, we present a RAG-based AI system benchmarking framework, named \pname{}, 
which allows users to: (1) easily configure the RAG pipeline via standard interfaces and examine the impact of different system configurations on the end-to-end performance and query quality; (2) model real-world scenarios via different parameters such as request rates, query/update ratio, batch sizes in its workload generator; 
and (3) automatically gather rich performance statistics at different stages of the RAG pipeline.
Specifically, \pname{} offers the following unique features.

\uline{First, \mbox{\pname{}} can emulate the dynamic behavior of real-world knowledge base operations via its workload generator (\S\ref{subsec:workoad-gen}).} By supporting various insert, update, and delete operations with configurable access patterns, \pname{} can model common scenarios such as continuously growing document collections, periodic content revisions, the removal of stale or invalid entries, as well as mixtures of these behaviors occurring concurrently. This flexibility allows \pname{} to capture realistic queries and to study how data freshness, indexing overhead, and end-to-end performance will be affected in diverse application scenarios.

\uline{Second, \mbox{\pname{}} can accurately capture the impact of changing system configurations in different pipeline stages (\S\ref{subsec:pipeline}).} It allows users to change the core parameters defined in the system, from embedding dimensions to the vector indexing method, 
from retrieval strategies to reranking algorithms. It systematically measures their impact on performance and query quality. With its profiling results, users can understand the performance behaviors after each configuration and tune system parameters accordingly, enabling data-driven decisions under diverse workload and deployment settings.

\uline{Third, \pname{} is designed to provide accurate measurements while keeping profiling overhead low (\S\ref{subsec:performance_metrics}). } It collects two complementary sets of metrics: performance metrics and accuracy metrics. The system performance metrics include fine-grained measurements of hardware resource utilization (e.g., GPU/CPU utilization, memory consumption, and I/O throughput) and end-to-end performance indicators (e.g., latency and throughput). The accuracy metrics evaluate the quality of generated answers, such as context recall, query accuracy, and factual consistency, and allow users to analyze the trade-offs between system efficiency and output quality. Together, these metrics provide a solid basis for understanding system behaviors of a RAG pipeline, which can guide its future deployment. \pname{} includes an adaptable system performance monitor that operates independently of the RAG pipeline, therefore, its performance profiling does not affect the regular RAG behaviors. \pname{} supports both system-level and component-level profiling, while introducing negligible runtime overhead (see our evaluation in $\S$\ref{subsec:overhead}).

\uline{Finally, \pname{} is implemented as a modular and extensible benchmarking framework with a well-defined interface (\S\ref{subsec:workflow}).} The RAG pipeline is decomposed into independent components for data embedding, indexing, retrieval, reranking, and generation, respectively. To support flexible pipeline construction, \pname{} defines only the input and output interface of each component, without imposing internal implementation constraints. Each component has its own configurable settings. By default, \pname{} supports multiple embedding models, rerank models, generation models, and vector databases, as presented in our evaluation ($\S$\ref{sec:workloads} and $\S$\ref{subsec:feature-eval}).


To reflect the performance behavior of real-world RAG systems, \pname{} collects a diverse set of datasets (Table~\ref{tab:retrieval_datasets}) that illustrate the heterogeneity of real-world content and cover different types of RAG applications. 
\pname{} provides a set of recent models across the entire RAG pipeline by default, covering embedding, reranking, and generation (Table~\ref{tab:llm_models}). It also supports popular vector databases like LanceDB, Milvus, Qdrant, Chroma, and Elasticsearch (Table~\ref{tab:vectordb_comparison}). 
We demonstrate the capability of \pname{} with a comprehensive set of experiments and show that \pname{} can conduct end-to-end performance breakdown for different RAG applications with minimal negative impact on the application performance. We also show that \pname{} can quantify the performance impact of various system configurations (e.g., available CPU cores, CPU memory, and GPU memory), as well as different RAG configurations (e.g., various batch sizes, embedding dimensions, and indexing schemes). We open source \pname{} at https://github.com/platformxlab/RAGPerf for public use. 




\section{Background and Motivation}
\label{sec:background}
In this section, we present the background of RAG-based AI systems. Then, we discuss the motivation for building an end-to-end benchmarking framework to examine performance and design trade-offs. 

\subsection{RAG-based AI System Architecture}
\label{subsec:rag-arch}
\begin{figure}
    \centering
    \includegraphics[width=0.55\linewidth]{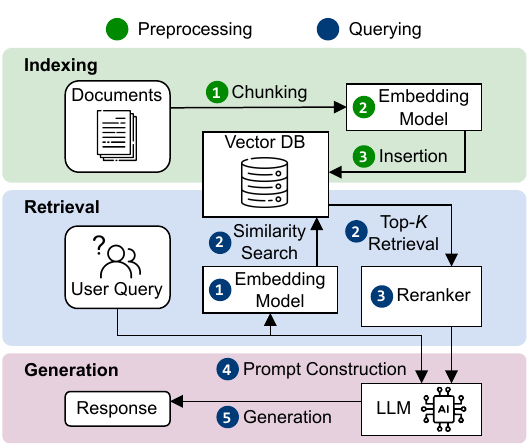}
    \vspace{-2ex}
    \caption{The end-to-end RAG workflow.}
    \label{fig:rag_workflow}
    \vspace{-3.5ex}
\end{figure}

LLMs are usually pre-trained and fine-tuned on publicly available datasets, making them powerful in general-purpose question answering. However, their effectiveness is limited in scenarios where user queries require domain-specific knowledge or private data to be answered. Retrieval-augmented generation (RAG)~\cite{karpukhin2020dense} has emerged as a method to supplement a language model with an external knowledge base. With RAG, LLMs can incorporate up-to-date knowledge, providing accurate responses in domain-specific scenarios (e.g., scientific, legal, and specific business). 

As shown in Figure~\ref{fig:rag_workflow}, a typical RAG system consists of three stages, where indexing serves as the knowledge base preparation while retrieval and generation occur when user queries are served. In the following, we explain the three stages, respectively.

\noindent
\underline{\textbf{Indexing.}}
To facilitate fast and accurate knowledge retrieval and provide LLMs with relevant context, the external knowledge base must be preprocessed to transform raw data into a searchable representation.
To be specific, the raw data are first segmented into smaller chunks (\circledg{1}), allowing fine-grained retrieval and minimizing the amount of irrelevant context introduced during subsequent retrieval and generation. Depending on the modality and structure of the input data as well as the requirements from downstream applications, the chunking may be performed at different granularities, ranging from text-based segmentation (e.g., fixed-length text segments, several sentences, or paragraphs) to modality-specific units such as image regions, video clips, or audio segments.
An embedding model is then used to map the resulting chunks into fixed-dimensional dense vectors (\circledg{2}). These vectors capture the semantic meaning of the chunks in a high-dimensional vector space. Within this space, semantically similar chunks are positioned closer to each other, providing the foundation for similarity-based retrieval.
The resulting vectors are then ingested into a vector database (\circledg{3}), where specialized indexing structures are constructed over the high-dimensional embedding space to enable efficient and scalable searching. Indexing methods fall into several families, ranging from hierarchical graph-based approaches (e.g., HNSW) to partition-based schemes (e.g., IVF). They sometimes incorporate compression techniques such as scalar quantization (SQ) and product quantization (PQ) to reduce index size and improve memory efficiency while preserving retrieval accuracy.

\noindent
\underline{\textbf{Retrieval.}}
Upon receiving user queries, the system first encodes user input prompts using the same embedding model employed during indexing (\circledb{1}). In practice, some RAG pipelines also introduce a lightweight auxiliary model that refines the search direction to further improve the retrieval accuracy. For example, the model may enhance the correspondence between the query representation and the retrieved content by decomposing or disambiguating the original query, or by generating refined query formulations. 
The resulting vector representations of the queries are then used to perform a similarity search to retrieve the top-$K$ most relevant context from the vector database (\circledb{2}). The similarity search is often based on approximate nearest-neighbor (ANN) algorithms, which trade off retrieval accuracy for greater computing efficiency, enabling low-latency search over large-scale, high-dimensional embedding spaces.
To further enhance retrieval precision, some pipelines incorporate a reranking step that re-evaluates the candidate contexts retrieved from the vector database (\circledb{3}). In this step, a more expressive model, often operating on the full query content and associated context, is used to reorder the candidate contexts and filter out less relevant or noisy results before passing the result to the downstream language model.

\noindent
\underline{\textbf{Generation.}} 
Finally, the RAG system combines the original user query and the retrieved contexts into a single prompt (\circledb{4}), typically following a pre-defined system template or user-defined prompt format that determines how the retrieved information is organized. The prompt is then passed to the language model, which uses both the query and the provided context to generate the final response (\circledb{5}).


\subsection{RAG-Based Applications}
\label{subsec:rag-demand}

\begin{table}[t]
    \centering
    \small
    \caption{Popular RAG-based applications.}
    \vspace{-2ex}
    \begin{tabular}{|>{\centering\arraybackslash}m{52pt}|>{\centering\arraybackslash}m{84pt}|>{\centering\arraybackslash}m{74pt}|}
    \hline
    \textbf{Category} & \textbf{Applications} & \textbf{Characteristics} \\ 
    \hline
    \multirow{4}{*}{\specialcell{Conversational \\ AI}}   & \specialcell{Search Engines \\ \cite{yan2024crag,cai2024forag,choi2024rac}}     & \multirow{4}{*}{\specialcell{High-concurrency user \\ requests that require \\short response time.}}      \\ \cline{2-2}
                                         & \specialcell{Customer Support \\ Bots \cite{xu2024retrieval,farnaz2026travquery,patel2025graph,gao2025olamind}}        &       \\ \cline{2-2}
                                         & \specialcell{Personal Assistants \cite{wang2024crafting,du2024perltqa}}   &       \\ \cline{2-2}
                                         \hline
    \multirow{3}{*}[-2ex]{\specialcell{Enterprise \\ Intelligence}}    & \specialcell{Legal Discovery \cite{barron2025bridging,de2025graph,buffa2025enhancing}}       &  \multirow{3}{*}[-1.5ex]{\specialcell{Structured Data with \\tables/forms. High \\precision and accuracy \\requirements.}}   \\ \cline{2-2}
                                                & \specialcell{Financial Report \\ Analysis \cite{choi2025finder,le2025rag,wang2025finsage}}  &    \\ \cline{2-2}
                                                & Medical Support \cite{wang2025mira,zhao2025medrag,xu2025mega}      &    \\ \cline{2-2}
                                            \hline
    \multirow{4}{*}{\specialcell{Multi-modal \\ Semantic Search}} & \specialcell{Video Retrieval \\ \cite{cao2025tvrag,mao2025multirag,luo2025videorag,jeong2025videorag,ren2025videorag}}      & \multirow{4}{*}{\specialcell{Heavy payload with \\ unstructured data.}}  \\ \cline{2-2}
                                                & \specialcell{Medical Imaging \\ Diagnosis \cite{ranjit2023rachestxray,song2025labrag,sun2025factawaremultimodalrag}}  &     \\ \cline{2-2}
                                                & \specialcell{Meeting Summary \cite{bhuvaji2025meetingrag,kang2025ragmeeting}}     &   \\ \cline{2-2}
    \hline
    \end{tabular}
    \vspace{-4ex}
\label{tab:applications}
\end{table}



RAG has evolved from an experimental technique into a critical infrastructure in different application domains. 
We summarize the popular RAG applications in Table~\ref{tab:applications}. 
The most prevalent application of RAG is general-purpose conversational AI, 
such as search engine-enhanced chatbots or real-time assistants. The system handles web-scale data ingestion while maintaining a short response time under high-concurrency user requests. The RAG pipeline needs to balance high-frequency vector updates to ensure knowledge freshness, while serving concurrent retrieval requests without latency degradation.

Another prevalent use case of RAG is to support grounded question answering in specialized enterprise domains such as finance, law, and medicine. 
Such enterprise RAG systems usually integrate an internal knowledge base containing complex, unstructured data. For instance, Financial Analysis systems extract fine-grained numerical data from earnings reports and tables; Legal Discovery tools have high-recall retrieval across vast contract repositories. Unlike open-domain AI applications, these applications must handle complex document layouts, such as multi-column PDFs and embedded tables.
In the meantime, these applications usually have strict privacy and high-precision standards. 

With the growth of multimedia content, RAG systems are also rapidly evolving to support cross-modal interaction beyond simple text. Typical examples include visual knowledge retrieval, video semantic search (e.g., retrieving relevant segments from videos), and audio analytics (e.g., meeting summary).
They require specialized data processing, such as optical character recognition (OCR)~\cite{tesseractocr,easyocr2020,rapidocr2021} or automatic speech recognition (ASR)~\cite{radford2022whisper,faster_whisper}. The RAG pipeline has a unified embedding strategy to map various file types into a common vector space.

These diverse RAG applications require drastically different resource requirements on the underlying hardware, and may experience different bottlenecks, such as memory capacity, I/O bandwidth, and compute capability.


\subsection{Benchmarks for RAG Systems}
\label{subsec:rag-bench}

\begin{table}[t!]
    \centering
    \footnotesize
    \caption{Existing benchmarks for RAG systems.}
    \vspace{-2ex}
    \begin{tabular}{|>{\centering\arraybackslash}m{56pt}|>{\centering\arraybackslash}m{44pt}|>{\centering\arraybackslash}m{25pt}|>{\centering\arraybackslash}m{30pt}|>{\centering\arraybackslash}m{30pt}|}
    \hline
    \textbf{Benchmarks} & \textbf{Configurable Pipeline} & \textbf{End-to-end} & \textbf{System Metrics} & \textbf{Semantic Metrics}  \\
    \hline
    BEIR~\cite{thakur2021beir}                      &    \rcross   &   \gcheck & \rcross & \gcheck \\
    RAGBench~\cite{friel2024ragbench}               &    \rcross   &   \gcheck & \rcross & \gcheck \\
    TruthfulQA~\cite{lin2022truthfulqa}             &    \rcross   &   \gcheck & \rcross & \gcheck \\
    RAGTruth~\cite{niu2024ragtruth}                 &    \rcross   &   \gcheck & \rcross & \gcheck \\
    FELM~\cite{zhao2023felm}                        &    \rcross   &   \gcheck & \rcross & \gcheck \\
    vectordbbench~\cite{vectordbbench}              &    \rcross   &   \rcross & \gcheck & \rcross \\
    vespa~\cite{vespa}                              &    \rcross   &   \rcross & \gcheck & \rcross \\
    LegalBench~\cite{pipitone2024legalbench}        &    \rcross   &   \gcheck & \rcross & \gcheck \\
    OmniEval~\cite{wang2025omnieval}                &    \rcross   &   \gcheck & \rcross & \gcheck \\
    AutoRAG~\cite{kim2024autorag}                   &    \gcheck   &   \gcheck & \rcross & \gcheck \\
    \hline
    \pname{}                                        &    \gcheck   &   \gcheck & \gcheck & \gcheck \\
    \hline
    \end{tabular}
\label{tab:benchmarks}
\vspace{-3ex}
\end{table}


Allocating resources across the components of a RAG pipeline involves numerous configuration choices. Understanding the performance implications of these decisions for a given application is important but challenging, as the interactions among multiple components and parameters make it difficult for developers to choose the optimal system configuration. Moreover, the complexity of integrating fragmented software components (e.g., data chunking, indexing, vector database, and LLM inference) places additional burdens on developers. As a result, studying the performance implications under realistic deployment scenarios will be hard and time-consuming without an end-to-end evaluation framework.

Unfortunately, existing RAG benchmarks fail to provide a holistic view of system behaviors, as shown in Table~\ref{tab:benchmarks}. A majority of current RAG benchmark frameworks prioritize the semantic metrics of the RAG application, including retrieval precision, hallucination rates, and generation faithfulness. While these metrics are crucial for ensuring output quality, they offer little insight into system efficiency (e.g., latency, throughput, and hardware utilization) in production deployment. Similarly, frameworks such as AutoRAG~\cite{kim2024autorag} focus on optimizing pipeline configurations based solely on response quality.
Other benchmarks such as vectordbbench~\cite{vectordbbench} and vespa~\cite{vespa} focus exclusively on individual components and only evaluate the vector databases or LLM generation speed. They fail to capture the runtime performance interference and resource contention in a fully integrated RAG pipeline. Furthermore, existing benchmarks generally lack the configurability. For example, LegalBench-RAG~\cite{pipitone2024legalbench} and OmniEval~\cite{wang2025omnieval} predefine their corpus, making it difficult for users to customize the RAG pipeline. For all these reasons, we are motivated to develop a unified and end-to-end benchmarking framework to ease the development and examination of RAG pipelines and applications. 



\section{Design and Implementation}
\label{sec:design}

\subsection{Goals of \pname{}}
\label{subsec:design-goal}
We develop \pname{} to meet the following goals. 

\begin{itemize}[leftmargin=*]
\item \textbf{End-to-end evaluation:} \pname{} should measure the performance behaviors and task quality of the entire RAG pipeline. Unlike component-wise benchmarks, \pname{} should be able to identify system-level performance bottlenecks. 
\item \textbf{Modularity and ease of use:} To accommodate the rapid evolution of RAG techniques, \pname{} should provide a fully modularized and configurable architecture. It should allow users to seamlessly integrate existing or customized retrievers, rerankers, and generators with minimal engineering effort.
\item \textbf{Low-overhead profiling:} \pname{} should provide a lightweight profiling method to collect fine-grained system metrics, e.g., memory footprints, GPU/CPU utilization,  I/O bandwidth, and others. 
\item \textbf{Workload diversity:} \pname{} should support realistic combinations of datasets, query types, input lengths, and arrival patterns. Diverse and configurable workloads can stress different aspects of the system under different operating conditions.
\item \textbf{Scalability and portability:} \pname{} should adapt to most platforms, whether deployed on a standalone GPU or multiple GPUs. The frameworks should also be able to deploy experimental settings and run experiments easily with different hardware settings.
\end{itemize}

\begin{figure}[t!]
    \centering
    \includegraphics[width=0.8\linewidth]{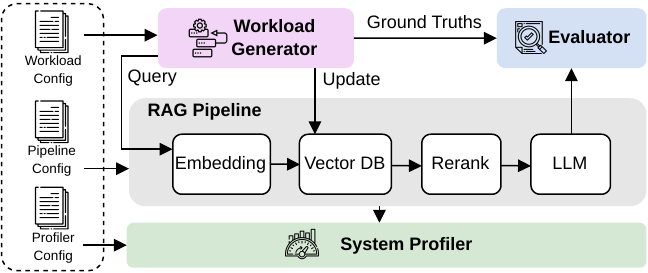}
    \vspace{-2ex}
    \caption{The architecture of our \pname{} framework.}
    \label{fig:rasbframework}
    \vspace{-3ex}
\end{figure}

\pname{} is designed to reflect real-world deployment while remaining easy for users to configure and use, as shown in Figure~\ref{fig:rasbframework}. We present the design and implementation of \pname{} as follows.

\subsection{Workload Generator}
\label{subsec:workoad-gen}



Real-world RAG systems often operate in environments where the underlying knowledge bases are continuously updated. Unlike indexing in traditional databases, vector indexing in RAG systems are more complicated and costly. First, each insertion requires transforming raw data into vectors using an embedding model, which is computationally expensive. Second, the support for updates and deletions is expensive, because many widely used indexing structures (e.g., HNSW~\cite{malkov2018hnsw} and IVF~\cite{sivic2003ivf}) do not support incremental updates and instead require costly index rebuilding. To mitigate the rebuilding overhead, many vector databases employ a small, temporary flat index to absorb incoming insertions and updates, trading a reduced rebuilding frequency for increased query latency. As this auxiliary index grows, indexing efficiency degrades, leading to higher query latency even though the total dataset size remains the same. Consequently, benchmarks that evaluate retrieval performance only on static datasets fail to capture the performance degradation introduced by continuous data updates in real-world RAG deployments. 

To simulate the real-world scenario, \pname{} generates concurrent read and write requests to stress the target RAG pipeline. 
A workload in \pname{} is composed of a sequence of four basic operations:
\begin{itemize}[leftmargin=*]
    \item Query: issues a user question from a predefined query pool to test data retrieval and generation.
    \item Insert: ingests entirely new data chunks, requiring the vector database to encode and index new embeddings.
    \item Update: modifies an existing document with new factual data, testing the ability to reflect changes in the knowledge base.
    \item Removal: deletes a specific file with file ID.
\end{itemize}
The overall composition of the workload is determined by a configuration with defined occurrence probability for each operation.

\noindent
\underline{Request Distribution.}
In addition to the mixture of operations, \pname{} also determines the access distribution, which controls how target file IDs are selected for updates and removals, as well as which documents are queried. \pname{} supports two primary access patterns to model different access behaviors. A \textit{Uniform} distribution ensures all files have an equal probability of being accessed or updated. \textit{Zipfian} distribution simulates a "hotspot", where a small subset of files receives the majority of updates and queries. 

By tuning these parameters, users can shape the workload to resemble specific deployment scenarios. For example, to compose a read-heavy workload with \textit{Wikipedia} dataset, we can generate a workload that has 90\% queries and 10\% updates alongside a Zipfian access distribution. The workload models real-world wiki behavior, where a small subset of trending wiki pages receives the vast majority of factual updates and queries, and most articles remain static.

\noindent
\underline{Request Generation.}
Once the operation and target file are selected based on distribution, \pname{} constructs the request payload. For \textit{insert}, \textit{removal}, and \textit{query} operations, required data are fetched directly from source dataset or question pool. However, the \textit{update} operation requires a more complex logic to ensure evaluation validity.

\begin{figure}[t!]
    \centering
    \includegraphics[width=0.9\linewidth]{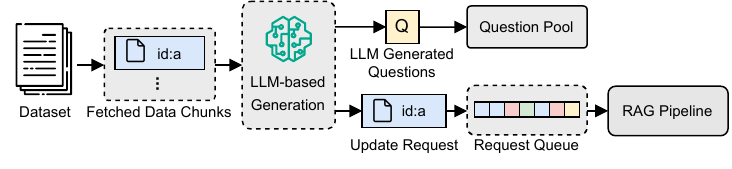}
    \vspace{-4ex}
    \caption{Update request generation in \pname{}.}
    \label{fig:rasbgenerator}
    \vspace{-3ex}
\end{figure}

\noindent
\underline{Dynamic Ground Truth Generation for Updates.}
To support accuracy and performance evaluation of RAG pipelines under update operations, \pname{} requires versioned data chunks as update payload and corresponding questions and ground-truth answers to test the updated knowledge base. Since most publicly available datasets do not support versioning, \pname{} uses an LLM-based generation module to generate both synthetic factual updates and their corresponding ground-truth question-answer pairs.

Given a user-defined update distribution, \pname{} samples a subset of documents for updating. For each document, \pname{} selects a single chunk at random and modifies a noun phrase or a numeric value. This ensures the update can be easily tested with a simple question. The modification is performed using DistilBERT~\cite{distilbert} model, where we mask the target words in the chunk and let the model generate replacement words at the mask position. The updated chunk is then packed as the request payload and issued to the RAG pipeline during benchmarking. The generator then uses the question generation model T5~\cite{t5} to read the updated data chunk and formulate a question whose answer is the replaced word. These questions are shuffled into the question pool to verify that the RAG system retrieves the updated fact rather than stale data.

Figure~\ref{fig:rasbgenerator} illustrates the generation of update request. The workload generator selects a list of target file IDs for updates by applying the user-defined access distribution across all file IDs. 
\pname{} then fetches the corresponding data chunks (the chunk with $id: a$ in Figure~\ref{fig:rasbgenerator}) from the dataset. All fetched data are passed to the LLM-based generation module to produce the updated chunks and test questions.
Finally, \pname{} builds the update requests with the modified chunk as the payload and dispatches them to the RAG pipeline during the benchmark run. The test questions are stored in the question pool and can be selected as a user query during the benchmark run.

\subsection{Configurable RAG Pipeline}
\label{subsec:pipeline}



A typical RAG system can expose tens to hundreds of configuration options and parameters. For instance, users may configure different vector database backends and further fine-tune their behaviors by choosing alternative indexing methods, quantization schemes, and similarity metrics. In addition, system-level parameters such as batch sizes, retrieval depth, and update strategies can significantly affect both performance and resource consumption. Together, these choices form a large and complex optimization space that is difficult to explore. 
Rather than treating the RAG pipeline as a monolithic black box, we decompose it into a set of modules whose behaviors are defined through external YAML configurations. 
At runtime, \pname{} collects fine-grained accuracy and performance metrics to quantify the performance and resource impact of different configurations.

\subsubsection{\bf Embedding}
\label{subsec:pipeline-embedding}

The performance and efficiency of the embedding stage depend on two critical factors: how raw data is chunked and how the embedding model is configured.

\noindent
\underline{Chunking.}
Embedding models typically operate under strict sequence length and assume that each basic chunk unit represents a semantically coherent piece of information. This makes chunking a critical part of data preprocessing before embedding.

For text-based documents, fixed-length chunking partitions documents into uniformly sized segments, offering a simple solution with predictable memory usage, but it may break semantic boundaries and can degrade retrieval relevance. Separator-based chunking leverages textual segments (e.g., sentences or paragraphs) to improve semantic coherence at the cost of having irregular embedding batch shapes and higher variance in access latency. Semantic-based chunking explicitly models content meaning to identify coherent boundaries. It typically uses NLP methods or simply a small language model to preserve semantically coherent context and reduce semantic fragmentation, thereby further improving retrieval accuracy and downstream generation quality at the cost of introducing larger computing overhead.

The overlap of chunks also affects the accuracy of data retrieval. In many cases, a segment of text may be contextually related to both its preceding and succeeding sentences, therefore, introducing overlaps helps preserve the shared context across adjacent chunks. 

\noindent
\underline{Document Format.}
Many knowledge sources used in RAG systems contain primarily textual information but are stored in presentation-oriented formats (e.g., PDF, Microsoft Word, and HTML). This introduces additional tradeoffs in the embedding pipeline.
One option is to extract only the text content of the document.
This method offers high flexibility with low computational overhead and enables the reuse of standard text embedding models, but may lose the structural or layout information, such as tables, figures, and formatting. An alternative approach is using OCR-based solutions, which offer a more reliable approach by directly operating on the rendered document layouts and can recover richer structural information. However, these methods incur substantially higher computational overhead and preprocessing latency. \pname{} supports both data chunking approaches.

\noindent
\underline{Embedding Model.}
After data chunking, we will encode each chunk into a single vector, which can capture the semantic meaning of the chunk in high-dimensional space.
Selecting an appropriate embedding model is a key design decision. Each model can be characterized by several key properties, including its parameter count, output embedding dimensionality, and computational cost. While larger models usually offer greater representational capacity, this typically comes at the expense of higher inference latency and increased resource consumption. 
The effectiveness of embedding is highly dependent on how the model is trained (including training dataset and training objectives). Models that are trained primarily on general conversational or natural-language corpora may perform poorly when applied to domain-specific inputs such as source code. Therefore, specialized embedding models tailored to particular domains are often required to achieve good retrieval quality.

Although many embedding models have relatively small memory footprints, colocating them with generation models on the GPU reduces the memory available for key–value (KV) caches for the generation model, which can degrade generation throughput and latency. Concurrent execution of embedding and generation workloads on the same device further introduces resource contention.  
Alternatively, offloading embedding computation to the host CPU reduces GPU memory pressure but incurs substantially higher embedding latency, due to the limited compute capability of CPUs. This may undermine the end-to-end response time despite reduced resource contention. These trade-offs highlight the need for careful resource allocation. \pname{} allows users to configure the hardware resource allocations (i.e., CPU/GPU and host/GPU memory) and examine their impact on the performance of RAG pipeline. 


\noindent
\underline{Multimodal Embedding.}
Instead of explicitly separating chunking and embedding, specialized multimedia or multimodal embedding models can operate directly on formatted inputs. 
These solutions offer a modular and easily integrable black-box approach, but limit the flexibility to tune chunking or embedding behavior independently. 
\pname{} has the option for users to use this approach. It provides multimodal embedding models such as ColPali~\cite{faysse2024colpali} and Clip~\cite{radford2021clip}. 

\noindent
\underline{Configuration and Profiling for Embedding.}
We support a wide range of input modalities through a modular and extensible preprocessing and embedding stack using \texttt{BaseEmbedder} abstraction (Figure~\ref{fig:rag_api}). For text-based documents, it incorporates popular open-source OCR pipelines and abstracts them into a unified module to facilitate easy adoption. 
At the embedding stage, \pname{} enables flexible integration of embedding models through the \texttt{SentenceTransformer} interface developed by HuggingFace, allowing users to evaluate different model architectures and embedding dimensions. \pname{} also provides prebuilt pipelines for audio, image, and video data, supporting both modality-to-text conversion and direct multimodal embedding to ensure consistent evaluation across diverse input formats.

\pname{} also collects auxiliary metadata at this stage to support performance analysis. During text chunking, \pname{} records the starting and ending offsets of each chunk, along with a reference to the original document, which incurs low storage overhead while enabling accurate tracing of chunking behavior. 
This helps examine the performance impact of the variance in chunk length within each input batch. During embedding, \pname{} reports data preparation overhead by measuring ingestion throughput and GPU memory utilization during batch encoding. \pname{} also quantifies the storage and memory costs of using higher-dimensional embedding models. 

\subsubsection{\bf Vector Database and Indexing.} 
\label{subsection:vector_database}
As the backbone of the RAG pipeline, the retrieval performance of the vector database is determined by three key factors that jointly affect indexing cost, insertion latency, query latency, and query quality.


\noindent
\underline{Indexing Method.} Graph-based indices (e.g., HNSW~\cite{malkov2018hnsw}) offer fast search speed, but require large memory capacity and long building time. Cluster-based indices (e.g., IVF~\cite{sivic2003ivf}) reduce memory footprints but require careful tuning of partition counts (\textit{nlist}) to balance search speed against training time. The parameters, such as the maximum graph degree (\textit{M}) and the size of the candidate set explored during construction (\textit{ef\_construction}), allow users to trade index building time for search performance and accuracy. 
To further reduce memory consumption, vector quantization is commonly applied to compress embedding representations. Popular techniques include scalar quantization (SQ), such as converting FP32 vectors to INT8, and product quantization (PQ), which compresses vectors by clustering. Using aggressive quantization substantially reduces memory usage and improves cache efficiency, but it inevitably introduces approximation error that 
degrades retrieval accuracy.

Vector databases have been offering GPU-accelerated indices to explore massive parallelisms for fast index construction and high-throughput retrieval. However, since RAG pipelines typically rely on the same GPUs for LLM generation, the vector database may contend GPU resources with the generation model.  

\noindent
\underline{Vector Database Update.}
To accommodate real-time updates without having an expensive rebuild of the full vector index, RAG systems often employ a hybrid indexing approach. It maintains a temporary flat index to cache incoming updates alongside the original approximate index. While this ensures immediate data freshness, it introduces a trade-off between query latency and rebuild overhead.

\begin{figure}[t!]
    \centering
    \includegraphics[width=0.99\linewidth]{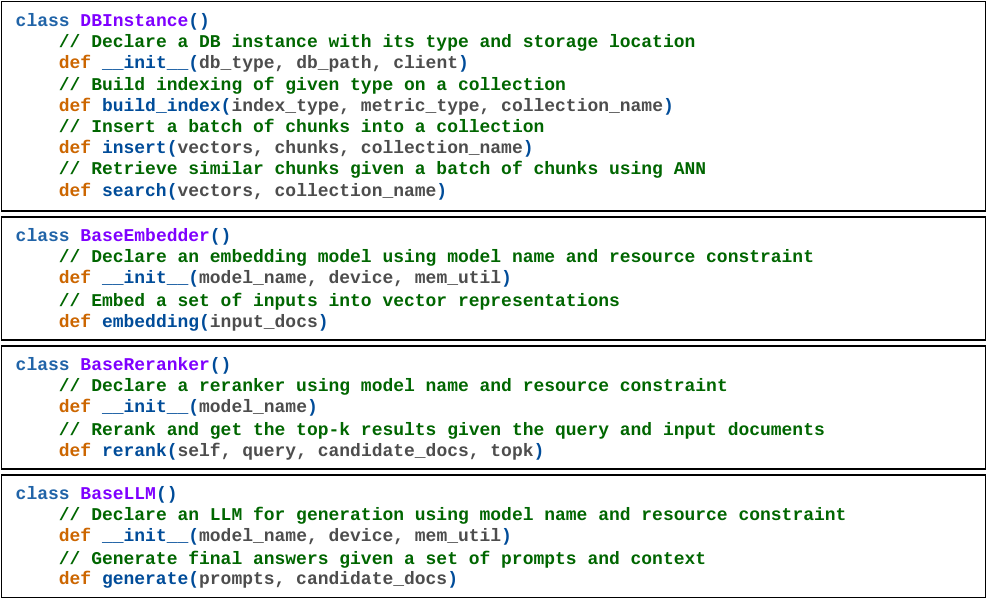}
    \vspace{-2ex}
    \caption{API for supporting different RAG pipeline modules. } 
    \label{fig:rag_api}
    \vspace{-4ex}
\end{figure}

\noindent
\underline{Configuration and Profiling for Vector Databases.}
The vector database ecosystem is highly fragmented with diverse APIs and deployment methods across providers (e.g., Milvus, Lance, and Qdrant). To support multiple vector databases in a single unified framework, \pname{} develops an \texttt{DBInstance} abstraction (Figure~\ref{fig:rag_api}) that defines a minimal set of standard operations. This abstraction decouples \pname{} from database-specific APIs, allowing new database backends to be integrated into the RAG pipeline, 
with a thin adapter layer that maps the standard interface to native database operations.

\pname{} tracks the execution time of core operations in vector databases, including insertion time, index building time, and query latency. If a hybrid index is enabled, \pname{} will report the latency for each index. It also monitors the storage footprint of the constructed index structure with a comparison to the raw vector data.
\pname{} records the retrieval result of each request in a compact binary format for the measurement of query accuracy. By storing only the retrieved chunk identifiers, 
\pname{} preserve the retrieval context for accuracy metrics, such as the recall (i.e., how often it successfully fetches the content needed to answer a query). 

\subsubsection{\bf Reranking.}
Reranking serves as an intermediate refinement step in RAG systems, with the goal of improving the quality of retrieved context through reordering and selective filtering.

\noindent
\underline{Rerank Model.} Selecting an appropriate reranking model is critical. Bi-encoder~\cite{devlin2019bertpretrainingdeepbidirectional} rerankers provide low-latency scoring and high throughput, it is suitable for latency-sensitive and large-scale deployment. In contrast, cross-encoders jointly process queries and documents to achieve higher ranking accuracy, but incur substantially higher computational cost. LLM-based rerankers can further improve ranking quality by using richer semantic reasoning, at the expense of significantly increased latency and resource consumption.

\noindent
\underline{Retrieval Depth.} Reranker effectiveness is strongly affected by the depth of the initial retrieval. Increasing the number of retrieved candidates increases the probability of including relevant content, but increases the computational cost of reranking. In addition, the selected output depth determines the amount of context forwarded to the generation model, directly affecting generation latency and quality.

\noindent
\underline{Configuration and Profiling for Reranking.}
Similar to the database abstraction, we develop a \texttt{BaseReranker} abstraction (Figure~\ref{fig:rag_api}) that standardizes the interface for various reranking strategies — from optimized cross-encoders (e.g., 
ms-marco-MiniLM~\cite{wang2020minilm}, BGE~\cite{xiao2024cpackpackedresourcesgeneral}, Cohere~\cite{cohere_rerank}) to LLM rankers (e.g., RankLLaMA~\cite{ma2023finetuningllamamultistagetext}, LLM4Rerank~\cite{gao2025llmrerank}). 
\pname{} records the reranked context to evaluate its effect by comparing accuracy metrics against retrieval. To reduce storage overhead, only chunk IDs are tracked rather than the full chunk content. 

\subsubsection{\bf Generation}
The generation step produces the final response by LLM inference using the user prompt and the retrieved context.

\noindent
\underline{Model Selection.} The choice of LLM significantly influences the reasoning capability and generation behavior of the RAG system.
Larger LLMs offer better response quality, while smaller models can provide lower latency and higher throughput. In addition, task-specific models, such as vision models, can be used to understand input contexts that contain images, tables, or structured documents.
\pname{} uses vLLM~\cite{kwon2023efficientmemorymanagementlarge}, an open-source high-throughput inference engine, as the default LLM backend. It supports a wide range of LLM architectures through a unified interface, allowing \pname{} to flexibly change models without additional changes to the rest of the pipeline.



\noindent
\underline{Parallelism.} \pname{} supports multiple parallelism strategies. Data parallelism replicates the model across devices and distributes incoming requests among replicas, improving throughput while keeping per-request latency the same. Tensor parallelism splits individual model layers across multiple devices, reducing per-layer computation time or allowing large models to fit in device memory. However, this approach requires high inter-GPU bandwidth (e.g., using NVLink~\cite{nvlink}). In comparison, pipeline parallelism fits large models by partitioning the model into stages that execute on different devices, but it introduces pipeline bubbles and can suffer from load imbalance.


\noindent
\underline{Configuration and Profiling for Generation.}
\pname{} uses vLLM as the backend serving framework for LLMs. It can retrieve detailed performance metrics, including time to first token (TTFT), time per output token (TPOT), and KV cache utilization, by querying the built-in metrics endpoint exposed by the vLLM. Users can analyze request queuing and scheduling delays through TTFT, decoding efficiency through TPOT, and memory pressure via KV cache utilization.




\subsection{Performance and Quality Metrics}
\label{subsec:performance_metrics}
\noindent
\underline{Performance Metrics.}
\pname{} integrates a resource monitor to help analyze the system resource usage. The monitor captures host-side CPU, memory, and I/O statistics via Linux \texttt{procfs}, GPU metrics through the GPU Performance Monitoring (GPM) interface in NVIDIA Management Library (NVML)~\cite{nvidia_nvml_2025}. 

All metrics are sampled at a configurable interval and recorded as time-series traces to capture resource utilization across all pipeline components. To minimize the interference with the RAG workload, we run the monitoring subsystem as a decoupled, low-priority background daemon. Profiled data is persisted to a separate storage device during execution, and the monitor tracks the time required to collect metrics from each probe and adjusts the sampling period automatically. The profiler supports graceful shutdown under crashes, ensuring all buffered data is safely flushed and persisted upon termination.


\noindent
\underline{Quality Metrics.}
\pname{} supports generation quality evaluation by integrating with \textit{Ragas}~\cite{es2025ragasautomatedevaluationretrieval} framework.  
It is an open-source framework that leverages LLMs as judges to score the RAG generation quality across multiple predefined metrics. \pname{} provides a wrapper to run Ragas locally via the vLLM engine and to pass the collected pipeline traces, including retrieved entries, generated responses, and ground-truth answers for evaluation. To avoid interfering with RAG workload, the evaluation is performed after the workload execution. 

\pname{} adopts three standard metrics to evaluate RAG generation quality.
\textit{Context Recall} evaluates retrieval effectiveness by computing the fraction of relevant chunks among all retrieved chunks from the vectorDB or after reranking.
\textit{Factual Consistency} evaluates the proportion of claims in the generated response that are explicitly supported by the retrieved context. A higher value means the LLM relies more heavily on the retrieved documents to generate its response, and vice versa.
\textit{Query Accuracy} measures the semantic alignment between the generated response and the ground-truth answer. 






\subsection{\pname{} Implementation}
\label{subsec:workflow}
\pname{} framework is implemented in Python and leverages existing open-source frameworks for module deployment as described below. 
\pname{} works directly with the HuggingFace ecosystem, enabling supported embedding models, language models, and datasets to be deployed via name. As for vector databases, \pname{} currently supports LanceDB~\cite{lancedb-multivector}, Milvus~\cite{milvus2021wang}, Qdrant~\cite{qdrant2025}, Chroma~\cite{chroma2025}, and Elasticsearch~\cite{elasticsearch2025}, using their open-sourced API that are publicly available. \pname{} uses vLLM as the LLM serving backend.

Different RAG pipeline components might have a different set of Python module requirements. To ensure reproducible and stable benchmarking environments, \pname{} handles the dependency resolution using \texttt{CMake} and \texttt{pip-tools}. The build system automatically resolves dependencies based on pipeline configuration, ensuring a consistent execution environment with easy deployment. \pname{} also supports isolating components that require incompatible dependencies in separate Python virtual environments or containers.

\pname{}'s monitoring subsystem is implemented as a multithreaded daemon that periodically samples system and component-level resource statistics from the Linux \texttt{proc} filesystem, cgroups v2, and NVIDIA GPU monitoring libraries. Host side system-wide metrics are collected from \texttt{/proc/}, while per-component statistics are obtained from \texttt{/proc/<pid>/}. For cgroup-level probing, the monitor reads from \texttt{/sys/fs/cgroup/<cg>/}. GPU metrics are collected with NVML GPM interface. We store output statistics with \texttt{protobuf} 30.2.


\section{Benchmark Workloads}
\label{sec:workloads}
We define a set of representative RAG workloads by default in \pname{}. We leverage these standardized workflows to stress-test the system, specifically targeting the variability in data modalities, model sizes, and retrieval logic. We detail their datasets, models, and vector database configurations in this section.

\subsection{Datasets}
\label{subsec:dataset}

\begin{table}[t]
    \centering
    \footnotesize
    \caption{Example datasets used in \pname{}. }
    \vspace{-2.5ex}
    \begin{tabular}{|>{\centering\arraybackslash}m{3cm}|>{\centering\arraybackslash}m{1cm}|>{\centering\arraybackslash}m{1cm}|>{\centering\arraybackslash}m{1cm}|N}
    \hline
    \textbf{Dataset} & \textbf{Type} & \textbf{Size} & \textbf{Entries} & \vspace{2pt}\quad \\
    \hline
    Wikipedia~\cite{wikidump}           & Text      & 19.3 GB   & 6.41M  \\
    Arxiv~\cite{kandpal2025commonpilev018tb}      & PDF       & 48 GB   & 30K  \\ 
    github-code~\cite{github-code}      & Code      & 32 GB   & 11M  \\
    The People’s Speech~\cite{galvez2021peoplesspeechlargescalediverse} & Audio & 35.5 GB & 0.3M \\
    \hline
    \end{tabular}
    \vspace{-2ex}
\label{tab:retrieval_datasets}
\end{table}

\pname{} uses various types of real-world datasets to model the knowledge sources for RAG systems in practice. As summarized in Table~\ref{tab:retrieval_datasets}, our selection covers different modalities and structural complexities. For general-purpose information retrieval systems, we use Wikipedia as the retrieval corpus paired with Natural Questions~\cite{2019naturalquestion} as queries. To address domain-specific challenges, we incorporate ArXiv research papers and GitHub repositories, using Machine Learning QA and coding assistant queries, respectively. Furthermore, we extend our evaluation to multi-modal contexts by including ArXiv page images and The People's Speech for vision and audio rag system support. To simulate deployment scenarios ranging from resource-constrained local devices to large-scale servers, we ensure all datasets support customizable scaling.


\begin{table}[t!]
\centering
\footnotesize
\caption{Models used for generation, embedding, and rerank.}
\vspace{-2.5ex}
{%
\begin{tabular}{|>{\centering\arraybackslash}m{4.7cm}|>{\centering\arraybackslash}m{2.2cm}|N}
    \hline
    \textbf{Generation Model}                               & \textbf{Tasks}    & \vspace{2pt}\quad \\
    \hline
    Qwen2.5-7B-Instruct~\cite{qwen2.5}                      & General \\
    gpt-oss-20b~\cite{openai2025gptoss120bgptoss20bmodel}   & General \\
    DeepSeek-R1-Distill-Qwen-32B~\cite{deepseekai2025}      & General \\
    Llama-3.3-70B-Instruct~\cite{touvron2023llama}          & General \\
    Qwen2.5-72B-Instruct                                    & General \\
    Qwen2.5-VL-3B-Instruct~\cite{qwen2.5vl}                 & Vision  \\
    Qwen2.5-VL-7B-Instruct                                  & Vision  \\
    Qwen2.5-VL-32B-Instruct                                 & Vision  \\
    Qwen2.5-Coder-7B-Instruct~\cite{qwen2.5-coder}          & Coding  \\
    Qwen2.5-Math-7B-Instruct~\cite{qwen2.5-math}            & Math    \\
    \hline
    \textbf{Embedding Model}                                & \textbf{Dimension} & \vspace{2pt}\quad \\
    \hline
    all-MiniLM-L6-v2~\cite{reimers-2019-sentence-bert}      & 384               \\
    all-mpnet-base-v2~\cite{reimers-2019-sentence-bert}     & 768               \\
    gte-large-en-v1.5~\cite{zhang2024mgte}                  & 1024              \\
    colpali-v1.2~\cite{faysse2024colpali}                   & 1030 $\times$ 128 \\ 
    clip-ViT-B-32~\cite{radford2021clip}                    & 512               \\ 
    \hline
    \textbf{Rerank Model}                                   & \textbf{Max Context} & \vspace{2pt}\quad \\
    \hline
    ms-marco-MiniLM ~\cite{wang2020minilm}                     & 512 tokens \\
    BGE~\cite{xiao2024cpackpackedresourcesgeneral}          & 8192 tokens \\
    \hline
\end{tabular}%
}
\label{tab:llm_models}
\vspace{-2ex}
\end{table}

\subsection{Models}
\label{subsec:models}


Our benchmark supports a comprehensive set of models across the entire RAG pipeline, covering embedding, reranking, and generation. In \pname{}, we aligning model capabilities with specific dataset in \S\ref{subsec:dataset} for benchmarking. For text generation, we employ models such as Qwen-2.5, Llama-3.3, and DeepSeek-R1 for general knowledge retrieval and reasoning. For domain-specific challenges, we integrate specialized models that include Qwen2.5-Coder for coding tasks, Qwen2.5-Math for mathematics, and Qwen2.5-VL for visual document understanding. To support these varied inputs during the retrieval stage, we utilize a range of embedding models, from standard dense retrievers like all-MiniLM and gte-large to the multi-modal ColPali model for visual PDF processing. We further refine retrieval precision by incorporating reranking models such as BGE and ms-marco-MiniLM, which accommodate varying context requirements. Across all stages, we include a diverse range of model sizes for generation and embedding model to evaluate the trade-offs between computational efficiency and output quality in diverse deployment environments. Table~\ref{tab:llm_models} details the supported models and their target tasks.


\subsection{Vector Databases}
\label{subsec:vectordb}

\pname{} supports LanceDB, Milvus, Qdrant, Chroma, and Elasticsearch as vector storage backends, together with multiple indexing methods, including HNSW, IVF, and DiskANN. The exact support matrix is summarized in Table~\ref{tab:vectordb_comparison}. Among these systems, LanceDB, Milvus, and Qdrant support GPU-based index construction, while Milvus and Qdrant additionally support GPU-accelerated querying.

\begin{table}[t!]
\centering
\footnotesize
\caption{Vector databases supported in \pname{}.}
\vspace{-2.5ex}
\begin{tabular}{|>{\centering\arraybackslash}m{1.85cm}|>{\centering\arraybackslash}m{3.5cm}|>{\centering\arraybackslash}m{2cm}|N}
\hline
\textbf{VectorDB}                                  & \textbf{Index Types}       & \textbf{Device Support} & \vspace{2pt}\quad \\
\hline
LanceDB~\cite{pace2025lanceefficientrandomaccess}  & IVF, HNSW, IVF-HNSW~\cite{baranchuk2018revisiting}                & CPU/GPU \\
Milvus~\cite{milvus2021wang}                       & HNSW, IVF, ScaNN~\cite{scann}, DiskANN~\cite{diskann}  & CPU/GPU \\
Qdrant~\cite{qdrant2025}                           & HNSW                       & CPU/GPU \\
Chroma~\cite{chroma2025}                           & HNSW                       & CPU \\
Elasticsearch~\cite{elasticsearch2025}             & HNSW, Flat                 & CPU \\
\hline
\end{tabular}
\label{tab:vectordb_comparison}
\vspace{-3ex}
\end{table}

\subsection{Multi-Modal Retrieval}
\label{subsec:multimodal}
To support different modalities, we implement separate pipelines for visual and audio data by default. 

\noindent\textbf{Visual Documents (PDF/Image).} \pname{} provide two distinct strategies for processing visual documents to address the trade-off between textual extraction accuracy and visual semantic preservation. The first approach is \textit{OCR-based extraction}, where we use Docling~\cite{livathinos2025doclingefficientopensourcetoolkit} to perform OCR. The OCR model extracts textual information or generates textual descriptions of visual elements and feeds them into text-based embeddings. 
Alternatively, we support \textit{visual embedding} by processing entire document pages directly, eliminating the need for OCR. We utilize ColPali~\cite{faysse2024colpali} to generate multi-vector embeddings for each page, employing a ColBERT-style~\cite{khattab2020colbertefficienteffectivepassage} reranker during retrieval to match query terms directly against visual page patches. Finally, the retrieved visual pages are fed into a vision-language model to generate the response.

\noindent\textbf{Audio.} For audio-based retrieval, we implement an ASR workflow. We utilize the Whisper~\cite{radford2022whisper} model to transcribe audio documents into text. Following transcription, the resulting text is processed using the standard text RAG pipeline, enabling effective retrieval of information contained within spoken content.

\begin{figure*}[t!]
    \begin{subfigure}[t]{0.55\linewidth}
    \centering
    \includegraphics[trim=0 1ex 0 0.5ex,clip,width=1\linewidth]{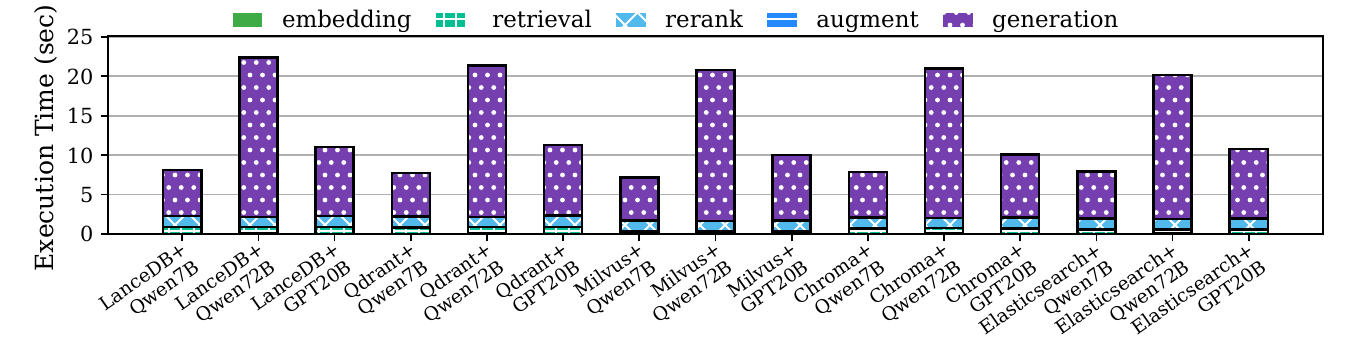}
    \vspace{-5ex}
    \caption{Text pipeline. }
    \label{fig:eval:textquerybreak}
    \end{subfigure}
    \begin{subfigure}[t]{0.35\linewidth}
    \centering
    \includegraphics[trim=0 0 0 0.5ex,clip,width=1\linewidth]{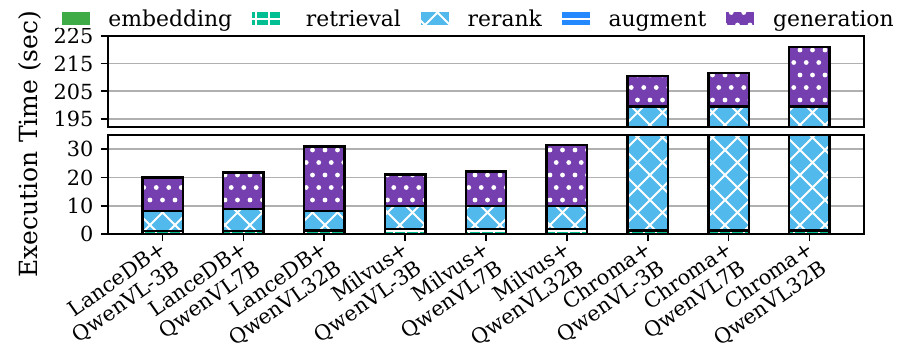}
    \vspace{-5ex}
    \caption{PDF/Image pipeline. }
    \label{fig:eval:pdfquerybreak}
    \end{subfigure}
    \vspace{-2.8ex}
    \caption{Execution time breakdown of a single batch of RAG application requests. }
    \label{fig:eval:querybreak}
    \vspace{-2ex}
\end{figure*}

\begin{figure*}[t!]
    \begin{subfigure}[t]{0.26\linewidth}
        \centering
        \includegraphics[trim=0 0ex 0 1.5ex,clip,width=\linewidth]{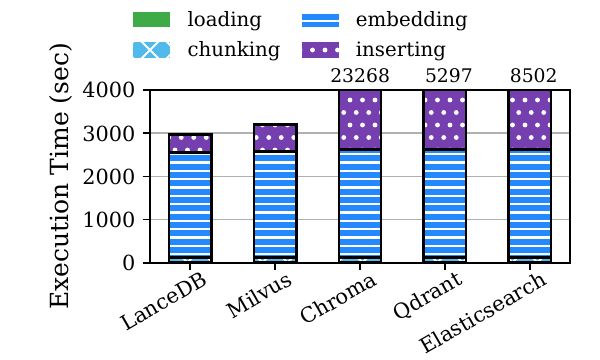}
        \vspace{-4ex}
        \caption{Text pipeline.}
        \label{fig:eval:textpreparebreak}
    \end{subfigure}
    \hspace{3ex}
    \begin{subfigure}[t]{0.29\linewidth}
        \centering
        \includegraphics[trim=0 0ex 0 1ex,clip,width=\linewidth]{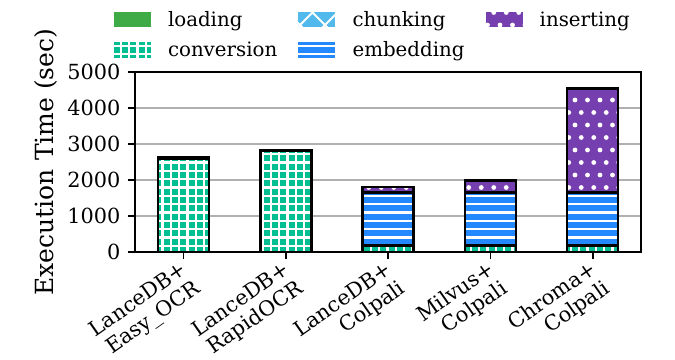}
        \vspace{-4ex}
        \caption{PDF/Image pipeline.}
        \label{fig:eval:pdfpreparebreak}
    \end{subfigure}
    \hspace{3ex}
    \begin{subfigure}[t]{0.32\linewidth}
        \centering
        \includegraphics[trim=0 0ex 0 1.5ex,clip,width=\linewidth]{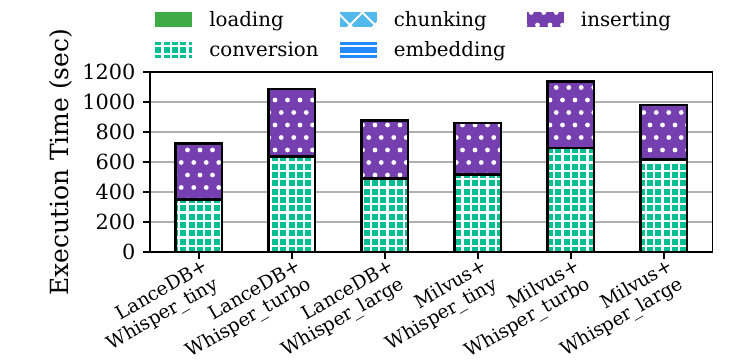}
        \vspace{-4ex}
        \caption{Audio pipeline.}
        \label{fig:eval:audiopreparebreak}
    \end{subfigure}
    \vspace{-2.8ex}
    \caption{Execution time breakdown of data indexing stage in RAG applications.}
    \label{fig:eval:preparebreak}
    \vspace{-2ex}
\end{figure*}

\section{Evaluation}
\label{subsec:feature-eval}
Our evaluation shows that \pname{} can (1) accurately profile the end-to-end performance of RAG applications (\S\ref{subsec:e2e-perf}); (2) identify performance bottlenecks through fine-grained hardware resource profiling (\S\ref{subsec:resource-util}); (3) evaluate RAG accuracy for different pipelines (\S\ref{subsec:accuracy});
(4) expose performance implications of update operations (\S\ref{subsec:update});
(5) quantify the impact of various system resource configurations on RAG performance (\S\ref{subsec:constraint});
and (6) conduct detailed sensitivity studies to isolate the impact of varying RAG parameters (\S\ref{subsec:sensitivity}). We also show that \pname{} incurs minimal overhead (\S\ref{subsec:overhead}).


\subsection{Experiment Setting}
\label{subsec:expr-setting}
\textbf{Testbed.}
Our testbed server is equipped with two 32-core AMD EPYC 9334 CPUs, 1.4TB DDR5 host memory, and two NVIDIA H100NVL GPUs with 94GB device memory. In \S\ref{subsec:e2e-perf} and \S\ref{subsec:resource-util}, we only use a single GPU for the experiments. 
We use Ubuntu 22.04 with the Linux kernel 5.15.0 and PyTorch 2.8.0+cu128. All datasets are stored in several 2TB Samsung 990 Pro SSDs. 


\noindent
\textbf{Dataset.}
We use the \textit{Wikipedia} dataset for the text pipeline. It consists of 6.41 million documents/articles totaling 19.3 GB of raw text. For the PDF/image pipeline, we select the 48GB \textit{Arxiv} dataset, which contains 30k PDF documents. We evaluate the audio retrieval using the \textit{PeoplesSpeech} dataset, comprising 35.5GB of audio data.

\subsection{End-to-end Performance}
\label{subsec:e2e-perf}
Figure~\ref{fig:eval:querybreak} breaks down the latency of the preprocessing and query stages for both the text-based and PDF/image-based RAG pipelines.

\noindent
\underline{Querying.}
Figure~\ref{fig:eval:textquerybreak} shows the latency breakdown of the text pipeline with a batch size of 64.
When using Qwen7B, GPT20B, and Qwen72B, the generation stage accounts for 75\%, 80\%, and 91\% of the total latency, respectively.
Consequently, the choice of vector database has a marginal performance impact on the end-to-end latency.

Figure~\ref{fig:eval:pdfquerybreak} presents the results of the PDF/image pipeline.
This pipeline is built based on ColPali's algorithm and exemplifies a wide range of PDF/image RAG pipelines.
In this pipeline, each reranking operation requires retrieving the complete source document, which incurs an average of 90 lookups to the vector database. As a result, reranking becomes costly, accounting for 28\% to 87\% of the total iteration time. 
The reranking phase dominates the total latency for Chroma, due to its suboptimal support for highly concurrent lookups.

\noindent
\underline{Indexing.}
Figure~\ref{fig:eval:preparebreak} shows a detailed latency breakdown of the indexing stage.
In the text pipeline (Figure~\ref{fig:eval:textpreparebreak}), the embedding stage duration is relatively stable, requiring 2450 seconds on average. 
In contrast, the insertion stage latency differs significantly across vector databases, with some (i.e., Chroma, Qdrant, and Elasticsearch) suffering longer insertion and index-building times. 
Specifically, Chroma suffers from a severe scalability bottleneck during insertion, increasing total execution time by 7.8$\times$ compared to LanceDB.

For PDF/image pipelines with OCR tools (e.g., EasyOCR~\cite{easyocr2020} and RapidOCR~\cite{rapidocr2021}), the format conversion dominates the execution time of the indexing stage, accounting for 98.2\% of the total duration on average. Although peak GPU utilization during OCR can be high, average utilization remains low (around 10\%), resulting in poor throughput even when multiple instances are deployed simultaneously. When using the ColPali model, we bypass text extraction entirely, shifting the cost to the embedding stage, where ColPali directly encodes document page images. On the vector database side, Chroma continues to exhibit low insertion throughput, reflecting the same scalability limitations observed during query processing.


Finally, the audio pipeline (Figure~\ref{fig:eval:audiopreparebreak}) is dominated by conversion and database insertion costs. The conversion time increases with model complexity. Whisper-turbo~\cite{radford2022whisper} requires approximately 612 seconds for transcription, 1.77$\times$ the time required by Whisper-tiny~\cite{radford2022whisper} (347 seconds). This increase is due to the higher computational cost. In addition, database insertion remains a significant overhead, accounting for up to 51\% of the total indexing time.

\insightbox{0.98\linewidth}{
    \begin{itemize}[leftmargin=*]
    \item \pname{} provides a detailed performance breakdown across modalities (text, PDF, audio), revealing distinct bottlenecks of each pipeline across various model and database configurations.
    \item For text query pipelines, LLM generation is the dominant bottleneck, making the choice of vector database less important, in terms of its impact on the end-to-end performance.
    \item For multimodal indexing, format conversion significantly contributes to the overall latency. The choice of conversion method significantly affects the indexing efficiency.
\end{itemize}}

\begin{figure*}[t!]
    \begin{subfigure}[t]{\linewidth}
        \centering
        \includegraphics[trim=0 0 0 2.5ex,clip,width=0.95\linewidth]{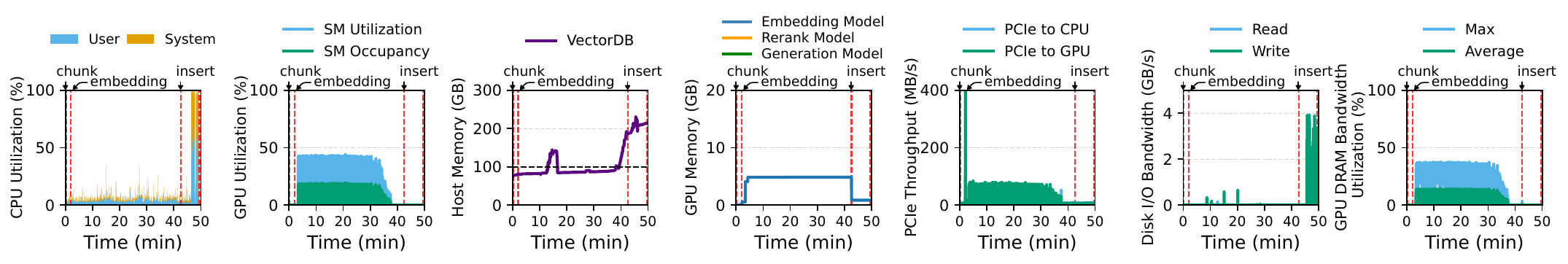}
        \vspace{-2.5ex}
        \caption{Data indexing stage for text pipeline. }
        \label{fig:eval:prepareutil}
    \end{subfigure}
    \begin{subfigure}[t]{\linewidth}
        \centering
        \includegraphics[trim=0 0 0 0,clip,width=0.935\linewidth]{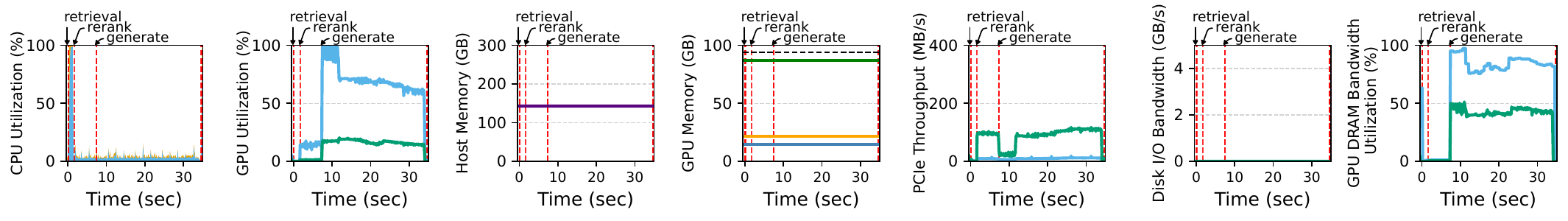}
        \vspace{-2.5ex}
        \caption{Single iteration in querying text pipeline. }
        \label{fig:eval:textutil}
    \end{subfigure}
    \begin{subfigure}[t]{1\linewidth}
        \centering
        \includegraphics[trim=0 0 0 0,clip,width=0.95\linewidth]{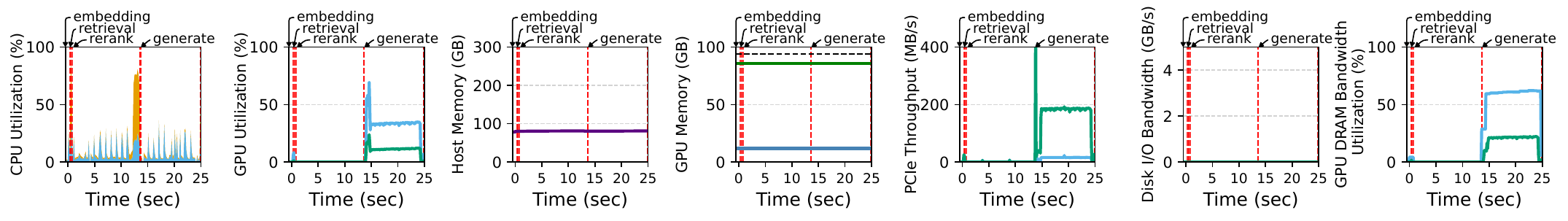}
        \vspace{-2.5ex}
        \caption{Single iteration in querying PDF pipeline. }
        \label{fig:eval:pdfutil}
    \end{subfigure}
    \vspace{-3.5ex}
    \caption{Resource utilization breakdown for different RAG applications. }
    \label{fig:eval:util}
    \vspace{-2ex}
\end{figure*}

\subsection{Resource Utilization Breakdown}
\label{subsec:resource-util}
A key capability of \pname{} is its fine-grained profiling of system resource utilization. Using the same workload in $\S$\ref{subsec:e2e-perf}, Figure~\ref{fig:eval:util} shows the resource utilization breakdown for indexing and query stages.

\noindent
\textbf{Compute Utilization.} As illustrated in Figure~\ref{fig:eval:util}, CPU utilization remains low (<5\%) during most stages, while GPU SM utilization dominates the embedding and generation stages, confirming that the core RAG logic is inference-bound. Moreover, we observe high SM utilization (60\%-70\% on average) paired with relatively low SM occupancy ($<15\%$), indicating that the number of active SMs is high even though the number of active warps relative to the hardware limit remains low. The reason is that the generation stage is memory-bandwidth bound rather than compute-bound, and it is further supported by the peak GPU DRAM bandwidth utilization observed during this window. Conversely, the CPU plays a critical role only during specific non-inference operations. CPU usage rises significantly during the insertion stage due to the intense overhead of building vector indices. The retrieval stage pushes CPU utilization to saturation at approximately 100\% (Figure~\ref{fig:eval:textutil}). Additionally, during the PDF rerank stage (Figure~\ref{fig:eval:pdfutil}), the CPU utilization spikes to 75\%, driven by the intense calculation of reranking scores for all retrieved chunks.

\noindent
\textbf{Memory Utilization.} GPU memory usage remains stable at approximately 90 GB throughout the query pipeline, dominated by the embedding model, reranking model, LLM weights, and key value cache. This behavior arises from the static memory allocation strategies used by modern serving backends such as vLLM, which keeps models resident even when inactive. While dynamic model offloading can reclaim unused memory, it introduces significant latency due to model reloading.
Moreover, host memory is mainly consumed by the vectorDB index, which becomes the primary bottleneck during vector retrieval and data processing. In the data indexing stage, insertion triggers a massive increase in host memory consumption, rising from around 90 GB to over 220 GB with the 20 GB Wikipedia dataset.
This indicates that update performance is fundamentally limited by available host memory.

\noindent
\textbf{I/O Traffic.}
The Disk I/O remains relatively idle during most stages of the RAG pipeline. However, the data indexing stage presents a major I/O bottleneck. As shown in Figure~\ref{fig:eval:prepareutil}, the insert stage demands a peak disk write throughput of 4 GB/s to persist the generated embeddings and vector indices to storage.


\insightbox{1\linewidth}{\begin{itemize}[leftmargin=*]
    \item \pname{} integrates a resource monitor that can capture fine-grained hardware utilization for each individual pipeline stage.
    \item The bottlenecks shift across different hardware resources at different RAG stages, with GPU during embedding and generation, CPU during retrieval and index building, and host memory during data processing, creating opportunities for system optimizations, such as overlapping pipeline executions.
    \item If all models in the pipeline require their weights to reside in GPU memory during execution, static resource allocation leaves memory underutilized when models are inactive. This calls for dynamic GPU memory scheduling across pipeline stages.
\end{itemize}}

\begin{figure}[t!]
    \centering
    \includegraphics[trim=0 0 0 1ex,clip,width=1\linewidth]{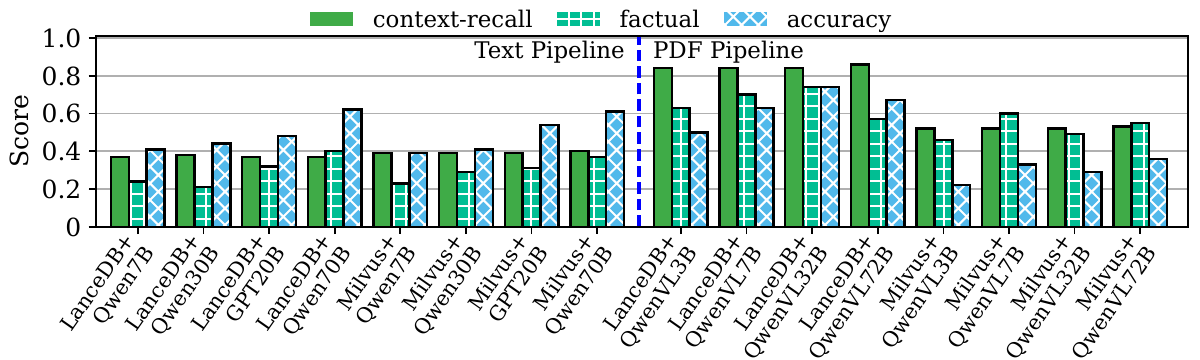}
    \vspace{-5ex}
    \caption{Accuracy score of Text and PDF pipeline.}
    \label{fig:eval:acc}
    \vspace{-2.5ex}
\end{figure}

\subsection{Accuracy Evaluation}
\label{subsec:accuracy}
Figure~\ref{fig:eval:acc} shows the accuracy for both text and PDF retrieval pipelines. We report the RAG quality using three metrics as defined in \S\ref{subsec:performance_metrics}: context-recall, factual consistency, and accuracy. The overall quality is primarily driven by the choice of the generation model rather than the vector database. As shown in the text pipeline in Figure~\ref{fig:eval:acc}, for the text pipeline, the context-recall scores remain nearly identical when swapping between LanceDB and Milvus under the same generation model. In contrast, scaling the generation model from Qwen-7B to Qwen-72B under the same LanceDB configuration improves the factual consistency score by $1.67\times$ and the accuracy score by $1.51\times$.


The PDF pipeline shows a different accuracy trend, driven by both retrieval quality and model capacity. Thanks to multi-vector retrieval and reranking~\cite{lancedb-multivector}, LanceDB configurations achieve an average context-recall of 0.84. With a sufficiently capable model such as QwenVL-32B, this high recall turns into a peak accuracy of 0.77. However, for small models, higher recall does not yield better accuracy. Milvus+QwenVL-3B achieves a context-recall of 0.52 but only produces an accuracy of 0.35, indicating that small vision-language models lack sufficient capacity to effectively utilize the retrieved PDF context despite its relevance. This suggests that in the PDF pipeline, model capacity acts as the bottleneck that determines whether high retrieval quality can be converted into accurate answers.

\insightbox{1\linewidth}{\begin{itemize}[leftmargin=*]
    \item A high context recall does not always guarantee better accuracy, as small models may lack the capacity to utilize retrieved context effectively.
\end{itemize}}


\begin{figure}[t!]
    \centering
    \includegraphics[trim=0 0 0 2.6ex,clip,width=0.95\linewidth]{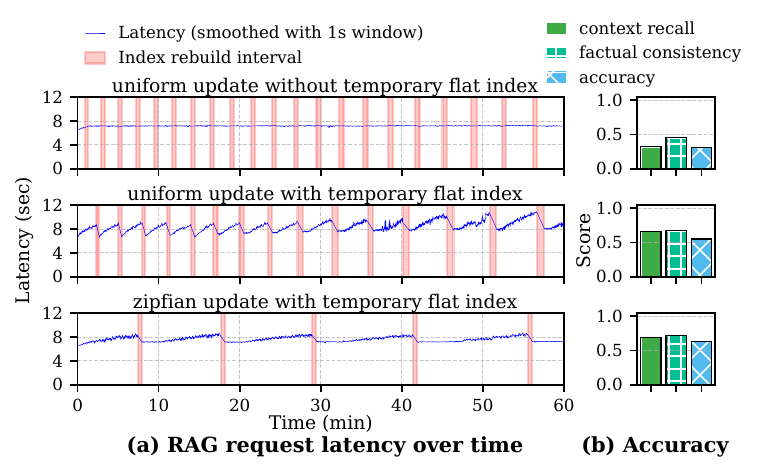}
    \vspace{-3ex}
    \caption{Latency of the text pipeline under update workload.}
    \label{fig:eval:update}
    \vspace{-3ex}
\end{figure}

\subsection{Update Operations}
\label{subsec:update}
We evaluate the text pipeline under continuous document updates using three configurations. We study how updates affect both request latency and response accuracy.
We configure \pname{} to issue a mixed workload (50\% queries and 50\% updates) against LanceDB using an IVF-HNSW~\cite{baranchuk2018revisiting} index over a 60-minute window with Qwen-7B as the generation model. Figure~\ref{fig:eval:update} shows the end-to-end latency of RAG requests over time with the corresponding accuracy score.

We first show the end-to-end performance without a temporary flat index configured under a uniform update distribution. Without the temporary flat index, newly inserted entries remain unsearchable until the next full index rebuild. Consequently, the latency remains stable at around 6.8 seconds, while the context recall and query accuracy remain low.

Enabling a temporary flat index in LanceDB resolves this by buffering new entries in a separate structure that is linearly scanned during each retrieval. This ensures that newly updated documents are immediately searchable, thereby restoring context recall and accuracy metrics. However, because the flat index grows between rebuilds, latency gradually increases from 7.01 to 10.7 seconds and drops sharply after each rebuild when the flat index is merged into the primary index.

In the third configuration, we apply the same temporary flat index but under a Zipfian update distribution, where modifications concentrate on a small subset of documents. The flat index accumulates fewer unique entries, resulting in a more gradual latency increase from 6.5 to 8.7 seconds and less frequent index rebuilds, while accuracy metrics remains the same.

\insightbox{1\linewidth}{\begin{itemize}[leftmargin=*]
    \item The dynamic workload generator in \pname{} enables users to evaluate the trade-off between retrieval quality and query performance across different update policies and distributions.
    \item A temporary flat index makes newly updated documents immediately searchable, improving retrieval quality at the cost of increased latency as the flat index grows. Search latency is stable if ignoring all updates before the next rebuild, but the pipeline retrieves stale data and degrades accuracy.
    \item With the temporary flat index, skewed update distributions result in lower latency overhead, as fewer unique entries are maintained in the flat structure compared to uniform updates.
\end{itemize}}

\subsection{Impact of Diverse Resource Configurations}
\label{subsec:constraint}
\begin{figure}[t!]
    \centering
    \includegraphics[trim=0 0 0 2ex,clip,width=0.95\linewidth]{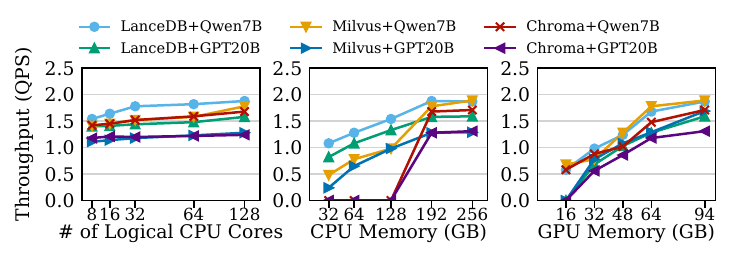}
    \vspace{-3.5ex}
    \caption{Text pipeline performance under limited resources.}
    \label{fig:eval:res}
    \vspace{-2.5ex}
\end{figure}

We now demonstrate \pname{}'s capability of quantifying the performance impact on RAG systems under different system resource configurations. We evaluate the performance of the RAG pipeline with different CPU, host memory, and GPU memory configurations. We present the end-to-end throughput (QPS) in Figure~\ref{fig:eval:res}.
First, CPU count does not significantly impact throughput. Compared to a system using 128 cores, the pipeline achieves 90.3\% of peak throughput at 32 cores and 78.2\% with 8 cores.
In contrast, host memory constraints impose a severe performance penalty. We configured Milvus and LanceDB to use disk-based vector indexing (DiskANN and IVF-HNSW), and under 32 GB available memory, throughput drops to 15.3\% and 37.6\% for Milvus and LanceDB, respectively. This degradation is primarily due to I/O overhead, which increases the retrieval latency by 6.1$\times$ to 12.5$\times$. Chroma, which relies exclusively on in-memory HNSW indexing, fails to execute when host memory is below 128 GB. Finally, GPU memory limitations significantly impact scalability. 
Reducing GPU memory capacity to 32 GB limits the inference batch size, decreasing average throughput to 47.1\%. All experiments with GPT-20B fail under the 16 GB GPU memory limit, as this capacity is insufficient to load the model weights. 


\insightbox{1\linewidth}{\begin{itemize}[leftmargin=*]
    \item The number of CPU cores has minimal impact on RAG throughput, as the retrieval and indexing algorithms are not compute-intensive. In contrast, insufficient host memory forces disk-based indexing, significantly degrading throughput.
    \item GPU memory is the primary hardware bottleneck, as it directly limits inference batch sizes and determines whether larger LLMs can be fully loaded into the GPU for serving.
\end{itemize}}

\begin{figure}[t!]
    \centering
    \includegraphics[trim=0 0 0 2ex,clip,width=0.85\linewidth]{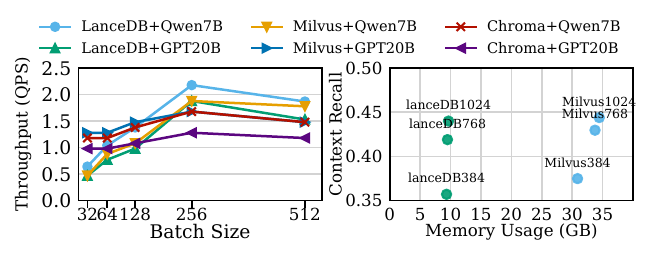}
    \vspace{-3.5ex}
    \caption{Text pipeline performance with various batch sizes and embedding dimensions.}
    \label{fig:eval:sensitivity}
    \vspace{-2ex}
\end{figure}

\begin{figure*}[t!]
    \centering
    \includegraphics[trim=0 0 0 2.8ex,clip,width=0.94\linewidth]{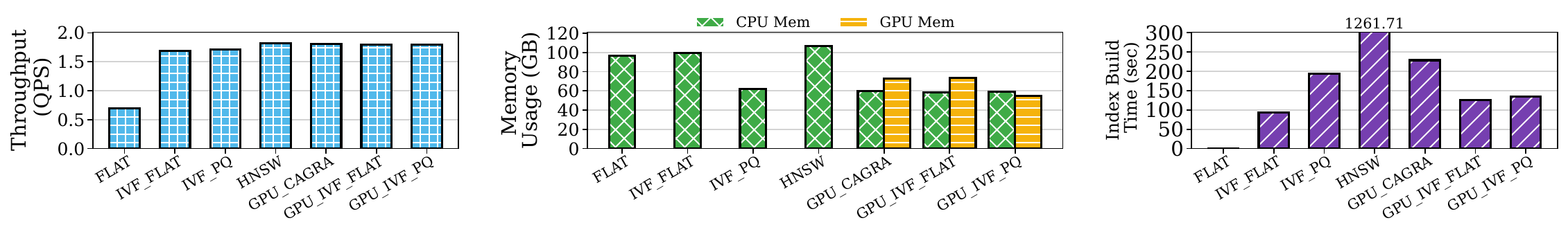}
    \vspace{-4ex}
    \caption{Text pipeline performance with different index schemes.}
    \label{fig:eval:textindex}
    \vspace{-3ex}
\end{figure*}

\subsection{Sensitivity Analysis}
\label{subsec:sensitivity}
\pname{} supports a wide range of pipeline settings (\S\ref{subsec:pipeline}). We demonstrate this by analyzing the end-to-end RAG throughput across a representative subset of parameters.

\noindent
\textbf{Batch Size.}
Figure~\ref{fig:eval:sensitivity} shows the performance impact of varying batch size. Increasing the batch size from 32 to 256 yields a significant throughput gain across all configurations. For example, the throughput of the LanceDB+Qwen7B configuration is improved by 3.6$\times$ due to increased GPU parallelism.
However, further increasing the batch size to 512 reduces throughput by up to 21\%, because the larger batch requires proportionally more GPU memory for KV cache, which limits the number of tokens decoded in parallel and forces sequential decoding steps, thereby increasing generation latency.

\noindent
\textbf{Embedding Dimension.}
We evaluate the trade-off between context recall and index memory usage across varying vector dimensions in Figure~\ref{fig:eval:sensitivity}. LanceDB exhibits significantly lower memory usage ($\approx$10~GB) compared to Milvus ($>$30~GB). This efficiency is attributed to LanceDB's architecture, which avoids loading the full index into host memory when opening a table; in contrast, Milvus loads the entire index into memory upon collection initialization. Notably, the increase in memory usage across different dimensions is minimal within the same database, with the IVF\_PQ index. This is because product quantization represents vectors primarily as fixed-length quantization codes, so increases in embedding dimension have little impact on overall index size. Regarding retrieval quality, higher-dimensional vectors consistently demonstrate better context recall.

\noindent
\textbf{Vector Database Indexing.} We present a comprehensive performance comparison of vectorDB index types in Figure~\ref{fig:eval:textindex} using Milvus as the DB platform, as it supports most vector indexing types.
The brute-force FLAT index serves as the baseline, limiting the end-to-end RAG throughput to 0.69 QPS. Other approximate nearest neighbor (ANN) indices achieve higher throughputs from 1.68 to 1.81 QPS. 
However, the resource requirements vary across different indexing methods: the HNSW index demands the highest cost, requiring over 100 GB of host memory and the longest build time ($>1200$s), making it unsuitable for update-heavy knowledge bases. In contrast, the quantized IVF\_PQ index offers the most effective balance, delivering strong throughput while being the fastest to build (200s) and highly memory-efficient.
Furthermore, our analysis suggests that GPU-based indices are less attractive, as their observed throughput gains are marginal ($1.04\times$) relative to their resource requirements ($70$GB GPU memory for GPU\_CAGRA), which is a key consideration given that the GPU device is also critical for the generation stage.

\insightbox{1\linewidth}{\begin{itemize}[leftmargin=*]
    \item Increasing batch size improves serving throughput. However, additional requests can take up GPU memory needed for the KV cache, resulting in overall performance reduction. 
    \item GPU-accelerated indices consume a large amount of GPU memory but offer only marginal performance gains. Instead, by reserving more GPU memory for LLM inference, we can deliver more performance benefits to the RAG pipeline.
\end{itemize}}


\subsection{Overhead Analysis}
\label{subsec:overhead}
We evaluate the overhead of \pname{} by measuring its interference with workloads and quantifying its resource consumption. We compare the end-to-end query latencies of both the text and the PDF pipeline with and without profiling enabled. The profiling has negligible overhead, increasing the single iteration time by only 0.11\%. Even at a sampling interval of 100 ms, the monitor generates only about 48 KB/s of disk write on average. To control memory overhead, \pname{} allocates a fixed-size circular buffer of 2 MB for each metric, preventing unbounded memory for long-running workloads. At the same time, buffering metrics in memory reduces the frequency of disk writes, thereby lowering CPU utilization. Overall, the profiler itself incurs minimal resource overhead, consuming less than 0.26\% of CPU usage and reaching a peak system DRAM usage of 243 MB.

\section{Related Work}

\noindent
\textbf{Benchmarking for RAG.}
Prior studies have proposed various benchmarks to evaluate RAG pipelines~\cite{thakur2021beir, friel2024ragbench, vectordbbench, vespa, kim2024autorag, krishna2025fact, lin2022truthfulqa, niu2024ragtruth, zhao2023felm, hsieh2024ruler, wang2025multimodal, wasserman2025real}. One category focuses on the semantic quality and accuracy of retrieved context and generated answers. For example, BEIR~\cite{thakur2021beir} and RAGBench~\cite{friel2024ragbench} provide diverse datasets to measure retrieval precision and generation faithfulness. Similarly, benchmarks such as TruthfulQA~\cite{lin2022truthfulqa}, RAGTruth~\cite{niu2024ragtruth}, and FELM~\cite{zhao2023felm} evaluate factuality, hallucination rates, and semantic alignment in RAG-generated answers. Another category evaluates the performance of individual RAG components. VectorDBBench~\cite{vectordbbench} and Vespa~\cite{vespa} evaluate the vector databases, but do not account for the generation stage. Frameworks like AutoRAG~\cite{kim2024autorag} and Frames~\cite{krishna2025fact} optimize pipeline configurations based on RAG response quality. 
Unlike prior studies which either limit their scope to accuracy metrics or evaluate the retrieval and generation separately, \pname{} introduces an end-to-end RAG-based AI system benchmarking framework.

\noindent
\textbf{System Optimization for RAG.}
Many studies have been conducted to optimize RAG system performance~\cite{jiang2025rago,mahapatra2025storage,jiang2024piperag,berchansky2023optimizing,jin2025ragcache}. RAGX~\cite{mahapatra2025storage} and IKS~\cite{quinn2025accelerating} improve retrieval performance by offloading ANN search to a near-storage accelerator. PipeRAG~\cite{jiang2024piperag} uses an algorithm-system co-design approach to improve system resource utilization. Berchansky et al.~\cite{berchansky2023optimizing} improve generation throughput by eliminating low-impact tokens in the retrieved context. System-oriented extensions further optimize RAG execution, such as caching reusable retrieval results to reduce latency and backend load~\cite{jin2025ragcache}. In alignment with these efforts, \pname{} offers an easy-to-use framework to facilitate system optimization with a detailed system profile and a comprehensive study of different design trade-offs in the RAG pipeline.

\noindent
\textbf{RAG Extension.}
Recent work extends the RAG paradigm with new retrieval and reasoning mechanisms~\cite{han2025graphrag,gao2023precise,vake2025bridging,zhang2025imprag,liu2025knowledge}.
HyDE~\cite{gao2023precise} and HyPE~\cite{vake2025bridging} improve the alignment between prompts and context by generating hypothetical documents or prompts. ImpRAG~\cite{zhang2025imprag} uses intermediate representations produced during generation as an embedding for retrieval. GraphRAG~\cite{han2025graphrag} extends standard RAG using knowledge graphs, while GGR~\cite{liu2025knowledge} extends it further by using graph neural networks (GNNs) to guide the retrieval. 
\pname{} can support such extensions thanks to its easy-configurable and modular design.

\section{Conclusion}
\label{sec:conclusion}

We develop \pname{}, an end-to-end benchmarking framework for characterizing the system behaviors of RAG pipelines.  
\pname{} includes the entire data embedding, indexing, retrieval, and generation workflow. It employs realistic multi-modal workload generators and system profiling to help developers identify the performance bottlenecks in the RAG pipeline. It also offers the flexibility for users to configure the RAG pipeline and explore different design trade-offs. Our experiments demonstrate that \pname{} can effectively quantify critical design trade-offs with negligible profiling overhead.
\balance



\newpage

\bibliographystyle{ACM-Reference-Format}
\bibliography{ref}

@article{friel2024ragbench,
  title={Ragbench: Explainable benchmark for retrieval-augmented generation systems},
  author={Friel, Robert and Belyi, Masha and Sanyal, Atindriyo},
  journal={arXiv preprint arXiv:2407.11005},
  year={2024}
}

@misc{distilbert,
      title={DistilBERT, a distilled version of BERT: smaller, faster, cheaper and lighter}, 
      author={Victor Sanh and Lysandre Debut and Julien Chaumond and Thomas Wolf},
      year={2020},
      eprint={1910.01108},
      archivePrefix={arXiv},
      primaryClass={cs.CL},
      url={https://arxiv.org/abs/1910.01108}, 
}

@misc{t5,
      title={Exploring the Limits of Transfer Learning with a Unified Text-to-Text Transformer}, 
      author={Colin Raffel and Noam Shazeer and Adam Roberts and Katherine Lee and Sharan Narang and Michael Matena and Yanqi Zhou and Wei Li and Peter J. Liu},
      year={2023},
      eprint={1910.10683},
      archivePrefix={arXiv},
      primaryClass={cs.LG},
      url={https://arxiv.org/abs/1910.10683}, 
}

@article{kim2024autorag,
  title={Autorag: Automated framework for optimization of retrieval augmented generation pipeline},
  author={Kim, Dongkyu and Kim, Byoungwook and Han, Donggeon and Eibich, Matou{\v{s}}},
  journal={arXiv preprint arXiv:2410.20878},
  year={2024}
}

@inproceedings{krishna2025fact,
  title={Fact, fetch, and reason: A unified evaluation of retrieval-augmented generation},
  author={Krishna, Satyapriya and Krishna, Kalpesh and Mohananey, Anhad and Schwarcz, Steven and Stambler, Adam and Upadhyay, Shyam and Faruqui, Manaal},
  booktitle={Proceedings of the 2025 Conference of the Nations of the Americas Chapter of the Association for Computational Linguistics: Human Language Technologies (Volume 1: Long Papers)},
  pages={4745--4759},
  year={2025}
}

@misc{thakur2021beir,
      title={BEIR: A Heterogenous Benchmark for Zero-shot Evaluation of Information Retrieval Models}, 
      author={Nandan Thakur and Nils Reimers and Andreas Rücklé and Abhishek Srivastava and Iryna Gurevych},
      year={2021},
      eprint={2104.08663},
      archivePrefix={arXiv},
      primaryClass={cs.IR},
      url={https://arxiv.org/abs/2104.08663}, 
}

@InProceedings{vespa,
    author="Paton, Norman W.
    and Williams, M. Howard
    and Dietrich, Kosmas
    and Liew, Olive
    and Dinn, Andrew
    and Patrick, Alan",
    editor="Lings, Brian
    and Jeffery, Keith",
    title="VESPA: A Benchmark for Vector Spatial Databases",
    booktitle="Advances in Databases",
    year="2000",
}

@misc{vectordbbench,
    author = {Zilliz},
    title = {zilliztech/VectorDBBench: A Benchmark Tool for VectorDB},
    year = {2024},
    howpublished = {\url{https://github.com/zilliztech/VectorDBBench}},
}

@inproceedings{lin2022truthfulqa,
      title={Truthfulqa: Measuring how models mimic human falsehoods},
      author={Lin, Stephanie and Hilton, Jacob and Evans, Owain},
      booktitle={Proceedings of the 60th annual meeting of the association for computational linguistics (volume 1: long papers)},
      pages={3214--3252},
      year={2022},
}

@inproceedings{niu2024ragtruth,
      title={Ragtruth: A hallucination corpus for developing trustworthy retrieval-augmented language models},
      author={Niu, Cheng and Wu, Yuanhao and Zhu, Juno and Xu, Siliang and Shum, Kashun and Zhong, Randy and Song, Juntong and Zhang, Tong},
      booktitle={Proceedings of the 62nd Annual Meeting of the Association for Computational Linguistics (Volume 1: Long Papers)},
      pages={10862--10878},
      year={2024}
}

@article{zhao2023felm,
      title={Felm: Benchmarking factuality evaluation of large language models},
      author={Zhao, Yiran and Zhang, Jinghan and Chern, I and Gao, Siyang and Liu, Pengfei and He, Junxian and others},
      journal={Advances in Neural Information Processing Systems},
      volume={36},
      pages={44502--44523},
      year={2023}
}

@article{hsieh2024ruler,
      title={RULER: What's the Real Context Size of Your Long-Context Language Models?},
      author={Hsieh, Cheng-Ping and Sun, Simeng and Kriman, Samuel and Acharya, Shantanu and Rekesh, Dima and Jia, Fei and Zhang, Yang and Ginsburg, Boris},
      journal={arXiv preprint arXiv:2404.06654},
      year={2024}
}

@inproceedings{wang2025multimodal,
      title={Multimodal needle in a haystack: Benchmarking long-context capability of multimodal large language models},
      author={Wang, Hengyi and Shi, Haizhou and Tan, Shiwei and Qin, Weiyi and Wang, Wenyuan and Zhang, Tunyu and Nambi, Akshay and Ganu, Tanuja and Wang, Hao},
      booktitle={Proceedings of the 2025 Conference of the Nations of the Americas Chapter of the Association for Computational Linguistics: Human Language Technologies (Volume 1: Long Papers)},
      pages={3221--3241},
      year={2025}
}

@article{wasserman2025real,
      title={REAL-MM-RAG: A Real-World Multi-Modal Retrieval Benchmark},
      author={Wasserman, Navve and Pony, Roi and Naparstek, Oshri and Goldfarb, Adi Raz and Schwartz, Eli and Barzelay, Udi and Karlinsky, Leonid},
      journal={arXiv preprint arXiv:2502.12342},
      year={2025}
}

@inproceedings{lewis2020rag,
  title={Retrieval-Augmented Generation for Knowledge-Intensive {NLP} Tasks},
  author={Lewis, Patrick and Perez, Ethan and Piktus, Aleksandra and Petroni, Fabio and Karpukhin, Vladimir and Goyal, Naman and K{\"u}ttler, Heinrich and Lewis, Mike and Yih, Wen-tau and Rockt{\"a}schel, Tim and others},
  booktitle={Advances in Neural Information Processing Systems (NeurIPS)},
  volume={33},
  pages={9459--9474},
  year={2020}
}

@inproceedings{karpukhin2020dense,
  title     = {Dense Passage Retrieval for Open-Domain Question Answering},
  author    = {Karpukhin, Vladimir and O{\u{g}}uz, Barlas and Min, Sewon and Lewis, Patrick and Wu, Ledell and Edunov, Sergey and Chen, Danqi and Yih, Wen-tau},
  booktitle = {Proceedings of the 2020 Conference on Empirical Methods in Natural Language Processing (EMNLP)},
  pages     = {6769--6781},
  year      = {2020},
  publisher = {Association for Computational Linguistics},
  doi       = {10.18653/v1/2020.emnlp-main.550},
  url       = {https://aclanthology.org/2020.emnlp-main.550}
}

@inproceedings{wang2025omnieval,
      title={Omnieval: An omnidirectional and automatic rag evaluation benchmark in financial domain},
      author={Wang, Shuting and Tan, Jiejun and Dou, Zhicheng and Wen, Ji-Rong},
      booktitle={Proceedings of the 2025 Conference on Empirical Methods in Natural Language Processing},
      pages={5737--5762},
      year={2025}
}

@article{pipitone2024legalbench,
      title={Legalbench-rag: A benchmark for retrieval-augmented generation in the legal domain},
      author={Pipitone, Nicholas and Alami, Ghita Houir},
      journal={arXiv preprint arXiv:2408.10343},
      year={2024}
}

@misc{wikidump,
    author = "Wikimedia Foundation",
    title  = "Wikimedia Downloads",
    year = 2025,
    url    = "https://dumps.wikimedia.org"
}

@misc{github-code,
    author = "codepattot",
    title  = "Github Code",
    year = 2026,
    url    = "https://huggingface.co/datasets/codeparrot/github-code"
}

@article{2019naturalquestion,
    author = {Kwiatkowski, Tom and Palomaki, Jennimaria and Redfield, Olivia and Collins, Michael and Parikh, Ankur and Alberti, Chris and Epstein, Danielle and Polosukhin, Illia and Devlin, Jacob and Lee, Kenton and Toutanova, Kristina and Jones, Llion and Kelcey, Matthew and Chang, Ming-Wei and Dai, Andrew
                        M. and Uszkoreit, Jakob and Le, Quoc and Petrov, Slav},
    title = {Natural Questions: A Benchmark for Question Answering
                    Research},
    journal = {Transactions of the Association for Computational Linguistics},
    volume = {7},
    pages = {453-466},
    year = {2019},
    month = {08},
    issn = {2307-387X},
    doi = {10.1162/tacl\_a\_00276},
    url = {https://doi.org/10.1162/tacl\_a\_00276},
    eprint = {https://direct.mit.edu/tacl/article-pdf/doi/10.1162/tacl\_a\_00276/1923288/tacl\_a\_00276.pdf},
}

@misc{kandpal2025commonpilev018tb,
      title={The Common Pile v0.1: An 8TB Dataset of Public Domain and Openly Licensed Text}, 
      author={Nikhil Kandpal and Brian Lester and Colin Raffel and Sebastian Majstorovic and Stella Biderman and Baber Abbasi and Luca Soldaini and Enrico Shippole and A. Feder Cooper and Aviya Skowron and John Kirchenbauer and Shayne Longpre and Lintang Sutawika and Alon Albalak and Zhenlin Xu and Guilherme Penedo and Loubna Ben Allal and Elie Bakouch and John David Pressman and Honglu Fan and Dashiell Stander and Guangyu Song and Aaron Gokaslan and Tom Goldstein and Brian R. Bartoldson and Bhavya Kailkhura and Tyler Murray},
      year={2025},
      eprint={2506.05209},
      archivePrefix={arXiv},
      primaryClass={cs.CL},
      url={https://arxiv.org/abs/2506.05209}, 
}

@misc{galvez2021peoplesspeechlargescalediverse,
      title={The People's Speech: A Large-Scale Diverse English Speech Recognition Dataset for Commercial Usage}, 
      author={Daniel Galvez and Greg Diamos and Juan Ciro and Juan Felipe Cerón and Keith Achorn and Anjali Gopi and David Kanter and Maximilian Lam and Mark Mazumder and Vijay Janapa Reddi},
      year={2021},
      eprint={2111.09344},
      archivePrefix={arXiv},
      primaryClass={cs.LG},
      url={https://arxiv.org/abs/2111.09344}, 
}

@misc{qwen2.5,
      title={Qwen2.5 Technical Report}, 
      author={Qwen and : and An Yang and Baosong Yang and Beichen Zhang and Binyuan Hui and Bo Zheng and Bowen Yu and Chengyuan Li and Dayiheng Liu and Fei Huang and Haoran Wei and Huan Lin and Jian Yang and Jianhong Tu and Jianwei Zhang and Jianxin Yang and Jiaxi Yang and Jingren Zhou and Junyang Lin and Kai Dang and Keming Lu and Keqin Bao and Kexin Yang and Le Yu and Mei Li and Mingfeng Xue and Pei Zhang and Qin Zhu and Rui Men and Runji Lin and Tianhao Li and Tianyi Tang and Tingyu Xia and Xingzhang Ren and Xuancheng Ren and Yang Fan and Yang Su and Yichang Zhang and Yu Wan and Yuqiong Liu and Zeyu Cui and Zhenru Zhang and Zihan Qiu},
      year={2025},
      eprint={2412.15115},
      archivePrefix={arXiv},
      primaryClass={cs.CL},
      url={https://arxiv.org/abs/2412.15115}, 
}

@misc{qwen2.5vl,
      title={Qwen2.5-VL Technical Report}, 
      author={Shuai Bai and Keqin Chen and Xuejing Liu and Jialin Wang and Wenbin Ge and Sibo Song and Kai Dang and Peng Wang and Shijie Wang and Jun Tang and Humen Zhong and Yuanzhi Zhu and Mingkun Yang and Zhaohai Li and Jianqiang Wan and Pengfei Wang and Wei Ding and Zheren Fu and Yiheng Xu and Jiabo Ye and Xi Zhang and Tianbao Xie and Zesen Cheng and Hang Zhang and Zhibo Yang and Haiyang Xu and Junyang Lin},
      year={2025},
      eprint={2502.13923},
      archivePrefix={arXiv},
      primaryClass={cs.CV},
      url={https://arxiv.org/abs/2502.13923}, 
}

@article{qwen2.5-coder,
      title={Qwen2. 5-Coder Technical Report},
      author={Hui, Binyuan and Yang, Jian and Cui, Zeyu and Yang, Jiaxi and Liu, Dayiheng and Zhang, Lei and Liu, Tianyu and Zhang, Jiajun and Yu, Bowen and Dang, Kai and others},
      journal={arXiv preprint arXiv:2409.12186},
      year={2024}
}

@article{qwen2.5-math,
  title={Qwen2.5-Math Technical Report: Toward Mathematical Expert Model via Self-Improvement}, 
  author={An Yang and Beichen Zhang and Binyuan Hui and Bofei Gao and Bowen Yu and Chengpeng Li and Dayiheng Liu and Jianhong Tu and Jingren Zhou and Junyang Lin and Keming Lu and Mingfeng Xue and Runji Lin and Tianyu Liu and Xingzhang Ren and Zhenru Zhang},
  journal={arXiv preprint arXiv:2409.12122},
  year={2024}
}

@misc{openai2025gptoss120bgptoss20bmodel,
      title={gpt-oss-120b \& gpt-oss-20b Model Card}, 
      author={OpenAI},
      year={2025},
      eprint={2508.10925},
      archivePrefix={arXiv},
      primaryClass={cs.CL},
      url={https://arxiv.org/abs/2508.10925}, 
}

@misc{deepseekai2025,
      title={DeepSeek-R1: Incentivizing Reasoning Capability in LLMs via Reinforcement Learning}, 
      author={DeepSeek-AI},
      year={2025},
      eprint={2501.12948},
      archivePrefix={arXiv},
      primaryClass={cs.CL},
      url={https://arxiv.org/abs/2501.12948}, 
}

@misc{touvron2023llama,
      title={LLaMA: Open and Efficient Foundation Language Models}, 
      author={Hugo Touvron and Thibaut Lavril and Gautier Izacard and Xavier Martinet and Marie-Anne Lachaux and Timothée Lacroix and Baptiste Rozière and Naman Goyal and Eric Hambro and Faisal Azhar and Aurelien Rodriguez and Armand Joulin and Edouard Grave and Guillaume Lample},
      year={2023},
      eprint={2302.13971},
      archivePrefix={arXiv},
      primaryClass={cs.CL},
      url={https://arxiv.org/abs/2302.13971}, 
}

@misc{faysse2024colpali,
  title={ColPali: Efficient Document Retrieval with Vision Language Models}, 
  author={Manuel Faysse and Hugues Sibille and Tony Wu and Bilel Omrani and Gautier Viaud and Céline Hudelot and Pierre Colombo},
  year={2024},
  eprint={2407.01449},
  archivePrefix={arXiv},
  primaryClass={cs.IR},
  url={https://arxiv.org/abs/2407.01449}, 
}

@inproceedings{radford2021clip,
  title={Learning transferable visual models from natural language supervision},
  author={Radford, Alec and Kim, Jong Wook and Hallacy, Chris and Ramesh, Aditya and Goh, Gabriel and Agarwal, Sandhini and Sastry, Girish and Askell, Amanda and Mishkin, Pamela and Clark, Jack and others},
  booktitle={International conference on machine learning},
  pages={8748--8763},
  year={2021},
  organization={PmLR}
}

@inproceedings{reimers-2019-sentence-bert,
  title     = {Sentence-BERT: Sentence Embeddings using Siamese BERT-Networks},
  author    = {Reimers, Nils and Gurevych, Iryna},
  booktitle = {Proceedings of the 2019 Conference on Empirical Methods in Natural Language Processing},
  year      = {2019},
  publisher = {Association for Computational Linguistics},
  url       = {https://arxiv.org/abs/1908.10084}
}

@article{zhang2024mgte,
  title={mGTE: Generalized Long-Context Text Representation and Reranking Models for Multilingual Text Retrieval},
  author={Zhang, Xin and Zhang, Yanzhao and Long, Dingkun and Xie, Wen and Dai, Ziqi and Tang, Jialong and Lin, Huan and Yang, Baosong and Xie, Pengjun and Huang, Fei and others},
  journal={arXiv preprint arXiv:2407.19669},
  year={2024}
}

@misc{pace2025lanceefficientrandomaccess,
      title={Lance: Efficient Random Access in Columnar Storage through Adaptive Structural Encodings}, 
      author={Weston Pace and Chang She and Lei Xu and Will Jones and Albert Lockett and Jun Wang and Raunak Shah},
      year={2025},
      eprint={2504.15247},
      archivePrefix={arXiv},
      primaryClass={cs.DB},
      url={https://arxiv.org/abs/2504.15247}, 
}

@inproceedings{milvus2021wang,
author = {Wang, Jianguo and Yi, Xiaomeng and Guo, Rentong and Jin, Hai and Xu, Peng and Li, Shengjun and Wang, Xiangyu and Guo, Xiangzhou and Li, Chengming and Xu, Xiaohai and Yu, Kun and Yuan, Yuxing and Zou, Yinghao and Long, Jiquan and Cai, Yudong and Li, Zhenxiang and Zhang, Zhifeng and Mo, Yihua and Gu, Jun and Jiang, Ruiyi and Wei, Yi and Xie, Charles},
title = {Milvus: A Purpose-Built Vector Data Management System},
year = {2021},
isbn = {9781450383431},
publisher = {Association for Computing Machinery},
address = {New York, NY, USA},
url = {https://doi.org/10.1145/3448016.3457550},
doi = {10.1145/3448016.3457550},
booktitle = {Proceedings of the 2021 International Conference on Management of Data},
pages = {2614–2627},
numpages = {14},
location = {Virtual Event, China},
series = {SIGMOD '21}
}

@misc{qdrant2025,
  author = {{Qdrant Team}},
  title = {Qdrant},
  year = {2025},
  url = {https://qdrant.tech/},
  note = {Vector Database}
}

@misc{chroma2025,
  author = {{Chroma Team}},
  title = {Chroma},
  year = {2025},
  url = {https://www.trychroma.com/},
  note = {Vector Database}
}

@misc{elasticsearch2025,
  author = {{Elastic}},
  title = {Elasticsearch},
  year = {2025},
  url = {https://www.elastic.co/elasticsearch/},
  note = {Distributed Search and Analytics Engine}
}

@misc{wang2020minilm,
      title={MiniLM: Deep Self-Attention Distillation for Task-Agnostic Compression of Pre-Trained Transformers}, 
      author={Wenhui Wang and Furu Wei and Li Dong and Hangbo Bao and Nan Yang and Ming Zhou},
      year={2020},
      eprint={2002.10957},
      archivePrefix={arXiv},
      primaryClass={cs.CL},
      url={https://arxiv.org/abs/2002.10957}, 
}

@misc{xiao2024cpackpackedresourcesgeneral,
      title={C-Pack: Packed Resources For General Chinese Embeddings}, 
      author={Shitao Xiao and Zheng Liu and Peitian Zhang and Niklas Muennighoff and Defu Lian and Jian-Yun Nie},
      year={2024},
      eprint={2309.07597},
      archivePrefix={arXiv},
      primaryClass={cs.CL},
      url={https://arxiv.org/abs/2309.07597}, 
}

@misc{lancedb-multivector,
  title = {Multivector Search in LanceDB},
  author = {{LanceDB Inc.}},
  year = {2025},
  howpublished = {\url{https://lancedb.com/docs/search/multivector-search/}},
  urldate = {2025-11-26},
  note = {LanceDB Documentation}
}

@misc{khattab2020colbertefficienteffectivepassage,
      title={ColBERT: Efficient and Effective Passage Search via Contextualized Late Interaction over BERT}, 
      author={Omar Khattab and Matei Zaharia},
      year={2020},
      eprint={2004.12832},
      archivePrefix={arXiv},
      primaryClass={cs.IR},
      url={https://arxiv.org/abs/2004.12832}, 
}

@misc{livathinos2025doclingefficientopensourcetoolkit,
      title={Docling: An Efficient Open-Source Toolkit for AI-driven Document Conversion}, 
      author={Nikolaos Livathinos and Christoph Auer and Maksym Lysak and Ahmed Nassar and Michele Dolfi and Panos Vagenas and Cesar Berrospi Ramis and Matteo Omenetti and Kasper Dinkla and Yusik Kim and Shubham Gupta and Rafael Teixeira de Lima and Valery Weber and Lucas Morin and Ingmar Meijer and Viktor Kuropiatnyk and Peter W. J. Staar},
      year={2025},
      eprint={2501.17887},
      archivePrefix={arXiv},
      primaryClass={cs.CL},
      url={https://arxiv.org/abs/2501.17887}, 
}

@misc{es2025ragasautomatedevaluationretrieval,
      title={Ragas: Automated Evaluation of Retrieval Augmented Generation}, 
      author={Shahul Es and Jithin James and Luis Espinosa-Anke and Steven Schockaert},
      year={2025},
      eprint={2309.15217},
      archivePrefix={arXiv},
      primaryClass={cs.CL},
      url={https://arxiv.org/abs/2309.15217}, 
}

@misc{nvidia_nvml_2025,
  author       = {NVIDIA},
  title        = {NVIDIA Management Library (NVML)},
  year         = {2025},
  howpublished = {\url{https://developer.nvidia.com/management-library-nvml}},
  note         = {Accessed: Augest 1st, 2025}
}

@article{oltp,
author = {Difallah, Djellel Eddine and Pavlo, Andrew and Curino, Carlo and Cudre-Mauroux, Philippe},
title = {OLTP-Bench: an extensible testbed for benchmarking relational databases},
year = {2013},
issue_date = {December 2013},
publisher = {VLDB Endowment},
volume = {7},
number = {4},
issn = {2150-8097},
url = {https://doi.org/10.14778/2732240.2732246},
doi = {10.14778/2732240.2732246},
month = dec,
pages = {277–288},
numpages = {12}
}

@inproceedings{ycsb,
author = {Cooper, Brian F. and Silberstein, Adam and Tam, Erwin and Ramakrishnan, Raghu and Sears, Russell},
title = {Benchmarking cloud serving systems with YCSB},
year = {2010},
isbn = {9781450300360},
publisher = {Association for Computing Machinery},
address = {New York, NY, USA},
url = {https://doi.org/10.1145/1807128.1807152},
doi = {10.1145/1807128.1807152},
booktitle = {Proceedings of the 1st ACM Symposium on Cloud Computing},
pages = {143–154},
numpages = {12},
location = {Indianapolis, Indiana, USA},
series = {SoCC '10}
}

@article{huang2024survey,
  title={A survey on retrieval-augmented text generation for large language models},
  author={Huang, Yizheng and Huang, Jimmy},
  journal={arXiv preprint arXiv:2404.10981},
  year={2024}
}

@article{gao2023retrieval,
  title={Retrieval-augmented generation for large language models: A survey},
  author={Gao, Yunfan and Xiong, Yun and Gao, Xinyu and Jia, Kangxiang and Pan, Jinliu and Bi, Yuxi and Dai, Yixin and Sun, Jiawei and Wang, Haofen and Wang, Haofen},
  journal={arXiv preprint arXiv:2312.10997},
  volume={2},
  number={1},
  year={2023}
}

@inproceedings{cai2024forag,
  title={Forag: Factuality-optimized retrieval augmented generation for web-enhanced long-form question answering},
  author={Cai, Tianchi and Tan, Zhiwen and Song, Xierui and Sun, Tao and Jiang, Jiyan and Xu, Yunqi and Zhang, Yinger and Gu, Jinjie},
  booktitle={Proceedings of the 30th ACM SIGKDD Conference on Knowledge Discovery and Data Mining},
  pages={199--210},
  year={2024}
}

@misc{yan2024crag,
  title        = {Corrective Retrieval Augmented Generation},
  author       = {Shi-Qi Yan and Jia-Chen Gu and Yun Zhu and Zhen-Hua Ling},
  year         = {2024},
  eprint       = {2401.15884},
  archivePrefix= {arXiv},
  primaryClass = {cs.CL}
}

@inproceedings{choi2024rac,
  title={RAC: Retrieval-augmented Conversation Dataset for Open-domain Question Answering in Conversational Settings},
  author={Choi, Bonggeun and Park, Jeongjae and Kim, Yoonsung and Park, Jaehyun and Ko, Youngjoong},
  booktitle={Proceedings of the 2024 Conference on Empirical Methods in Natural Language Processing: Industry Track},
  pages={1477--1488},
  year={2024}
}

@inproceedings{xu2024retrieval,
  title={Retrieval-augmented generation with knowledge graphs for customer service question answering},
  author={Xu, Zhentao and Cruz, Mark Jerome and Guevara, Matthew and Wang, Tie and Deshpande, Manasi and Wang, Xiaofeng and Li, Zheng},
  booktitle={Proceedings of the 47th international ACM SIGIR conference on research and development in information retrieval},
  pages={2905--2909},
  year={2024}
}

@inproceedings{farnaz2026travquery,
    author="Farnaz, Aneela and Huyck, Chris",
    editor="Bramer, Max and Stahl, Frederic",
    title="TravQuery: A Customer Support Chatbot Based on Retrieval Augmented Generation (RAG)",
    booktitle="Artificial Intelligence XLII",
    year="2026",
    publisher="Springer Nature Switzerland",
    address="Cham",
    pages="130--140",
    isbn="978-3-032-11402-0"
}

@article{patel2025graph,
  title={Graph-Enhanced Retrieval-Augmented Question Answering for E-Commerce Customer Support},
  author={Patel, Piyushkumar},
  journal={arXiv preprint arXiv:2509.14267},
  year={2025}
}

@article{gao2025olamind,
  title={OlaMind: Towards Human-Like and Hallucination-Safe Customer Service for Retrieval-Augmented Dialogue},
  author={Gao, Tianhong and Shen, Jundong and Shi, Bei and Wang, Jiapeng and Ju, Ying and Yao, Junfeng and Ran, Jiao and Zhang, Yong and Dong, Lin and Yu, Huiyu and others},
  journal={arXiv preprint arXiv:2510.22143},
  year={2025}
}

@article{wang2024crafting,
  title={Crafting personalized agents through retrieval-augmented generation on editable memory graphs},
  author={Wang, Zheng and Li, Zhongyang and Jiang, Zeren and Tu, Dandan and Shi, Wei},
  journal={arXiv preprint arXiv:2409.19401},
  year={2024}
}

@article{du2024perltqa,
  title={Perltqa: A personal long-term memory dataset for memory classification, retrieval, and synthesis in question answering},
  author={Du, Yiming and Wang, Hongru and Zhao, Zhengyi and Liang, Bin and Wang, Baojun and Zhong, Wanjun and Wang, Zezhong and Wong, Kam-Fai},
  journal={arXiv preprint arXiv:2402.16288},
  year={2024}
}

@article{barron2025bridging,
  title={Bridging Legal Knowledge and AI: Retrieval-Augmented Generation with Vector Stores, Knowledge Graphs, and Hierarchical Non-negative Matrix Factorization},
  author={Barron, Ryan C and Eren, Maksim E and Serafimova, Olga M and Matuszek, Cynthia and Alexandrov, Boian S},
  journal={arXiv preprint arXiv:2502.20364},
  year={2025}
}

@article{de2025graph,
  title={Graph RAG for Legal Norms: A Hierarchical and Temporal Approach},
  author={de Martim, Hudson},
  journal={arXiv preprint arXiv:2505.00039},
  year={2025}
}

@article{buffa2025enhancing,
  title={Enhancing legal document building with Retrieval-Augmented Generation},
  author={Buffa, Matteo and Ferrara, Alfio and Picascia, Sergio and Riva, Davide and Castano, Silvana},
  journal={Computer Law \& Security Review},
  volume={59},
  pages={106229},
  year={2025},
  publisher={Elsevier}
}

@inproceedings{choi2025finder,
  title={Finder: Financial dataset for question answering and evaluating retrieval-augmented generation},
  author={Choi, Chanyeol and Kwon, Jihoon and Ha, Jaeseon and Choi, Hojun and Kim, Chaewoon and Lee, Yongjae and Sohn, Jy-yong and Lopez-Lira, Alejandro},
  booktitle={Proceedings of the 6th ACM International Conference on AI in Finance},
  pages={638--646},
  year={2025}
}

@techreport{le2025rag,
  title={RAG-IT: Retrieval-Augmented Instruction Tuning for Automated Financial Analysis},
  author={Le, Van-Duc and To, Hai-Thien},
  year={2025},
  institution={arXiv. org}
}

@inproceedings{wang2025finsage,
  title={Finsage: A multi-aspect rag system for financial filings question answering},
  author={Wang, Xinyu and Chi, Jijun and Tai, Zhenghan and Kwok, Tung Sum Thomas and He, Hailin and Li, Zhuhong and Hua, Yuchen and Li, Muzhi and Lu, Peng and Wang, Suyucheng and others},
  booktitle={Proceedings of the 34th ACM International Conference on Information and Knowledge Management},
  pages={6144--6152},
  year={2025}
}

@inproceedings{wang2025mira,
    author = {Wang, Jinhong and Ashraf, Tajamul and Han, Zongyan and Laaksonen, Jorma and Anwer, Rao Muhammad},
    title = {MIRA: A Novel Framework for Fusing Modalities in Medical RAG},
    year = {2025},
    isbn = {9798400720352},
    publisher = {Association for Computing Machinery},
    address = {New York, NY, USA},
    url = {https://doi.org/10.1145/3746027.3755760},
    doi = {10.1145/3746027.3755760},
    booktitle = {Proceedings of the 33rd ACM International Conference on Multimedia},
    pages = {6307–6315},
    numpages = {9},
    location = {Dublin, Ireland},
    series = {MM '25}
}

@inproceedings{zhao2025medrag,
    title={Medrag: Enhancing retrieval-augmented generation with knowledge graph-elicited reasoning for healthcare copilot},
    author={Zhao, Xuejiao and Liu, Siyan and Yang, Su-Yin and Miao, Chunyan},
    booktitle={Proceedings of the ACM on Web Conference 2025},
    pages={4442--4457},
    year={2025}
}

@article{xu2025mega,
    title={MEGA-RAG: a retrieval-augmented generation framework with multi-evidence guided answer refinement for mitigating hallucinations of LLMs in public health},
    author={Xu, Shan and Yan, Zhaokun and Dai, Chengxiao and Wu, Fan},
    journal={Frontiers in Public Health},
    volume={13},
    pages={1635381},
    year={2025},
    publisher={Frontiers Media SA}
}

@inproceedings{cao2025tvrag,
    author = {Cao, Zongsheng and He, Yangfan and Liu, Anran and Xie, Jun and Chen, Feng and Wang, Zhepeng},
    title = {TV-RAG: A Temporal-aware and Semantic Entropy-Weighted Framework for Long Video Retrieval and Understanding},
    year = {2025},
    isbn = {9798400720352},
    publisher = {Association for Computing Machinery},
    address = {New York, NY, USA},
    url = {https://doi.org/10.1145/3746027.3755873},
    doi = {10.1145/3746027.3755873},
    booktitle = {Proceedings of the 33rd ACM International Conference on Multimedia},
    pages = {9071–9079},
    numpages = {9},
    location = {Dublin, Ireland},
    series = {MM '25}
}

@misc{mao2025multirag,
      title={Multi-RAG: A Multimodal Retrieval-Augmented Generation System for Adaptive Video Understanding}, 
      author={Mingyang Mao and Mariela M. Perez-Cabarcas and Utteja Kallakuri and Nicholas R. Waytowich and Xiaomin Lin and Tinoosh Mohsenin},
      year={2025},
      eprint={2505.23990},
      archivePrefix={arXiv},
      primaryClass={cs.AI},
      url={https://arxiv.org/abs/2505.23990}, 
}

@misc{luo2025videorag,
    title={Video-RAG: Visually-aligned Retrieval-Augmented Long Video Comprehension}, 
    author={Yongdong Luo and Xiawu Zheng and Guilin Li and Shukang Yin and Haojia Lin and Chaoyou Fu and Jinfa Huang and Jiayi Ji and Fei Chao and Jiebo Luo and Rongrong Ji},
    year={2025},
    eprint={2411.13093},
    archivePrefix={arXiv},
    primaryClass={cs.CV},
    url={https://arxiv.org/abs/2411.13093}, 
}

@misc{jeong2025videorag,
    title={VideoRAG: Retrieval-Augmented Generation over Video Corpus}, 
    author={Soyeong Jeong and Kangsan Kim and Jinheon Baek and Sung Ju Hwang},
    year={2025},
    eprint={2501.05874},
    archivePrefix={arXiv},
    primaryClass={cs.CV},
    url={https://arxiv.org/abs/2501.05874}, 
}

@misc{ren2025videorag,
    title={VideoRAG: Retrieval-Augmented Generation with Extreme Long-Context Videos}, 
    author={Xubin Ren and Lingrui Xu and Long Xia and Shuaiqiang Wang and Dawei Yin and Chao Huang},
    year={2025},
    eprint={2502.01549},
    archivePrefix={arXiv},
    primaryClass={cs.IR},
    url={https://arxiv.org/abs/2502.01549}, 
}

@misc{ranjit2023rachestxray,
      title={Retrieval Augmented Chest X-Ray Report Generation using OpenAI GPT models}, 
      author={Mercy Ranjit and Gopinath Ganapathy and Ranjit Manuel and Tanuja Ganu},
      year={2023},
      eprint={2305.03660},
      archivePrefix={arXiv},
      primaryClass={cs.CL},
      url={https://arxiv.org/abs/2305.03660}, 
}

@misc{song2025labrag,
      title={LaB-RAG: Label Boosted Retrieval Augmented Generation for Radiology Report Generation}, 
      author={Steven Song and Anirudh Subramanyam and Irene Madejski and Robert L. Grossman},
      year={2025},
      eprint={2411.16523},
      archivePrefix={arXiv},
      primaryClass={cs.CV},
      url={https://arxiv.org/abs/2411.16523}, 
}

@misc{sun2025factawaremultimodalrag,
      title={Fact-Aware Multimodal Retrieval Augmentation for Accurate Medical Radiology Report Generation}, 
      author={Liwen Sun and James Zhao and Megan Han and Chenyan Xiong},
      year={2025},
      eprint={2407.15268},
      archivePrefix={arXiv},
      primaryClass={cs.CL},
      url={https://arxiv.org/abs/2407.15268}, 
}

@inproceedings{bhuvaji2025meetingrag,
    author = {Bhuvaji, Sartaj and Chouhan, Prachitee and Irukulla, Madhuroopa and Singhvi, Jay and Bae, Wan D. and Alkobaisi, Shayma},
    title = {A Retrieval-Augmented Framework For Meeting Insight Extraction},
    year = {2025},
    isbn = {9798400706295},
    publisher = {Association for Computing Machinery},
    address = {New York, NY, USA},
    url = {https://doi.org/10.1145/3672608.3707915},
    doi = {10.1145/3672608.3707915},
    booktitle = {Proceedings of the 40th ACM/SIGAPP Symposium on Applied Computing},
    pages = {899–906},
    numpages = {8},
    location = {Catania International Airport, Catania, Italy},
    series = {SAC '25}
}

@misc{kang2025ragmeeting,
      title={GETALP@AutoMin 2025: Leveraging RAG to Answer Questions based on Meeting Transcripts}, 
      author={Jeongwoo Kang and Markarit Vartampetian and Felix Herron and Yongxin Zhou and Diandra Fabre and Gabriela Gonzalez-Saez},
      year={2025},
      eprint={2508.00476},
      archivePrefix={arXiv},
      primaryClass={cs.CL},
      url={https://arxiv.org/abs/2508.00476}, 
}

@misc{nvlink,
    title        = {NVIDIA NVLink and NVLink Switch},
    year         = {2026},
    author       = {{NVIDIA}},
    howpublished = {\url={https://www.nvidia.com/en-us/data-center/nvlink/}}, 
}

@inproceedings{jiang2025rago,
    title={Rago: Systematic performance optimization for retrieval-augmented generation serving},
    author={Jiang, Wenqi and Subramanian, Suvinay and Graves, Cat and Alonso, Gustavo and Yazdanbakhsh, Amir and Dadu, Vidushi},
    booktitle={Proceedings of the 52nd Annual International Symposium on Computer Architecture},
    pages={974--989},
    year={2025}
}

@inproceedings{mahapatra2025storage,
    title={In-Storage Acceleration of Retrieval Augmented Generation as a Service},
    author={Mahapatra, Rohan and Santhanam, Harsha and Priebe, Christopher and Xu, Hanyang and Esmaeilzadeh, Hadi},
    booktitle={Proceedings of the 52nd Annual International Symposium on Computer Architecture},
    pages={450--466},
    year={2025}
}

@article{jiang2024piperag,
    title={Piperag: Fast retrieval-augmented generation via algorithm-system co-design},
    author={Jiang, Wenqi and Zhang, Shuai and Han, Boran and Wang, Jie and Wang, Bernie and Kraska, Tim},
    journal={arXiv preprint arXiv:2403.05676},
    year={2024}
}

@article{berchansky2023optimizing,
    title={Optimizing retrieval-augmented reader models via token elimination},
    author={Berchansky, Moshe and Izsak, Peter and Caciularu, Avi and Dagan, Ido and Wasserblat, Moshe},
    journal={arXiv preprint arXiv:2310.13682},
    year={2023}
}

@inproceedings{quinn2025accelerating,
  title={Accelerating retrieval-augmented generation},
  author={Quinn, Derrick and Nouri, Mohammad and Patel, Neel and Salihu, John and Salemi, Alireza and Lee, Sukhan and Zamani, Hamed and Alian, Mohammad},
  booktitle={Proceedings of the 30th ACM International Conference on Architectural Support for Programming Languages and Operating Systems, Volume 1},
  pages={15--32},
  year={2025}
}

@article{jin2025ragcache,
  title={Ragcache: Efficient knowledge caching for retrieval-augmented generation},
  author={Jin, Chao and Zhang, Zili and Jiang, Xuanlin and Liu, Fangyue and Liu, Shufan and Liu, Xuanzhe and Jin, Xin},
  journal={ACM Transactions on Computer Systems},
  volume={44},
  number={1},
  pages={1--27},
  year={2025},
  publisher={ACM New York, NY}
}

@inproceedings{liu2025knowledge,
  title={Knowledge Graph Retrieval-Augmented Generation via GNN-Guided Prompting},
  author={Liu, Haochen and Wang, Song and Li, Jundong},
  booktitle={Second Conference on Language Modeling},
  year={2025}
}

@article{zhang2025imprag,
  title={ImpRAG: Retrieval-Augmented Generation with Implicit Queries},
  author={Zhang, Wenzheng and Lin, Xi Victoria and Stratos, Karl and Yih, Wen-tau and Chen, Mingda},
  journal={arXiv preprint arXiv:2506.02279},
  year={2025}
}

@inproceedings{gao2023precise,
  title={Precise zero-shot dense retrieval without relevance labels},
  author={Gao, Luyu and Ma, Xueguang and Lin, Jimmy and Callan, Jamie},
  booktitle={Proceedings of the 61st Annual Meeting of the Association for Computational Linguistics (Volume 1: Long Papers)},
  pages={1762--1777},
  year={2023}
}

@article{vake2025bridging,
  title={Bridging the question-answer gap in retrieval-augmented generation: Hypothetical prompt embeddings},
  author={Vake, Domen and Vi{\v{c}}i{\v{c}}, Jernej and To{\v{s}}i{\'c}, Aleksandar},
  journal={IEEE access},
  year={2025},
  publisher={IEEE}
}

@misc{han2025graphrag,
      title={Retrieval-Augmented Generation with Graphs (GraphRAG)}, 
      author={Haoyu Han and Yu Wang and Harry Shomer and Kai Guo and Jiayuan Ding and Yongjia Lei and Mahantesh Halappanavar and Ryan A. Rossi and Subhabrata Mukherjee and Xianfeng Tang and Qi He and Zhigang Hua and Bo Long and Tong Zhao and Neil Shah and Amin Javari and Yinglong Xia and Jiliang Tang},
      year={2025},
      eprint={2501.00309},
      archivePrefix={arXiv},
      primaryClass={cs.IR},
      url={https://arxiv.org/abs/2501.00309}, 
}

@article{malkov2018hnsw,
  title={Efficient and robust approximate nearest neighbor search using hierarchical navigable small world graphs},
  author={Malkov, Yu A and Yashunin, Dmitry A},
  journal={IEEE transactions on pattern analysis and machine intelligence},
  volume={42},
  number={4},
  pages={824--836},
  year={2018},
  publisher={IEEE}
}

@inproceedings{sivic2003ivf,
  title={Video Google: A text retrieval approach to object matching in videos},
  author={Sivic and Zisserman},
  booktitle={Proceedings ninth IEEE international conference on computer vision},
  pages={1470--1477},
  year={2003},
  organization={IEEE}
}

@misc{devlin2019bertpretrainingdeepbidirectional,
      title={BERT: Pre-training of Deep Bidirectional Transformers for Language Understanding}, 
      author={Jacob Devlin and Ming-Wei Chang and Kenton Lee and Kristina Toutanova},
      year={2019},
      eprint={1810.04805},
      archivePrefix={arXiv},
      primaryClass={cs.CL},
      url={https://arxiv.org/abs/1810.04805}, 
}

@INPROCEEDINGS{tesseractocr,
  author={Smith, R.},
  booktitle={Ninth International Conference on Document Analysis and Recognition (ICDAR 2007)}, 
  title={An Overview of the Tesseract OCR Engine}, 
  year={2007},
  volume={2},
  number={},
  pages={629-633},
  doi={10.1109/ICDAR.2007.4376991}}

@misc{easyocr2020,
  title        = {{EasyOCR}: Ready-to-use {OCR} with 80+ Supported Languages},
  author       = {{JaidedAI}},
  year         = {2020},
  howpublished = {\url{https://github.com/JaidedAI/EasyOCR}}
}

@misc{rapidocr2021,
  title        = {{RapidOCR}: {OCR} Toolbox},
  author       = {{RapidAI Team}},
  year         = {2021},
  howpublished = {\url{https://github.com/RapidAI/RapidOCR}}
}

@misc{radford2022whisper,
      title={Robust Speech Recognition via Large-Scale Weak Supervision}, 
      author={Alec Radford and Jong Wook Kim and Tao Xu and Greg Brockman and Christine McLeavey and Ilya Sutskever},
      year={2022},
      eprint={2212.04356},
      archivePrefix={arXiv},
      primaryClass={eess.AS},
      url={https://arxiv.org/abs/2212.04356}, 
}

@software{faster_whisper,
      title={faster-whisper},
      author={{SYSTRAN}},
      year={2023},
      url={https://github.com/SYSTRAN/faster-whisper},
      note={Reimplementation of OpenAI's Whisper using CTranslate2}
}

@misc{ma2023finetuningllamamultistagetext,
      title={Fine-Tuning LLaMA for Multi-Stage Text Retrieval}, 
      author={Xueguang Ma and Liang Wang and Nan Yang and Furu Wei and Jimmy Lin},
      year={2023},
      eprint={2310.08319},
      archivePrefix={arXiv},
      primaryClass={cs.IR},
      url={https://arxiv.org/abs/2310.08319}, 
}

@inproceedings{gao2025llmrerank,
title={{LLM}4Rerank: {LLM}-based Auto-Reranking Framework for Recommendations},
author={Jingtong Gao and Bo Chen and Xiangyu Zhao and Weiwen Liu and Xiangyang Li and Yichao Wang and Wanyu Wang and Huifeng Guo and Ruiming Tang},
booktitle={THE WEB CONFERENCE 2025},
year={2025},
url={https://openreview.net/forum?id=HEBVEmK22u}
}

@misc{cohere_rerank,
  author = {Cohere},
  title = {Cohere Rerank},
  year = {2023},
  howpublished = {\url{https://docs.cohere.com/docs/rerank-overview}},
}

@misc{kwon2023efficientmemorymanagementlarge,
      title={Efficient Memory Management for Large Language Model Serving with PagedAttention}, 
      author={Woosuk Kwon and Zhuohan Li and Siyuan Zhuang and Ying Sheng and Lianmin Zheng and Cody Hao Yu and Joseph E. Gonzalez and Hao Zhang and Ion Stoica},
      year={2023},
      eprint={2309.06180},
      archivePrefix={arXiv},
      primaryClass={cs.LG},
      url={https://arxiv.org/abs/2309.06180}, 
}

@inproceedings{scann,
  title={Accelerating Large-Scale Inference with Anisotropic Vector Quantization},
  author={Guo, Ruiqi and Sun, Philip and Lindgren, Erik and Geng, Quan and Simcha, David and Chern, Felix and Kumar, Sanjiv},
  booktitle={International Conference on Machine Learning},
  year={2020},
  URL={https://arxiv.org/abs/1908.10396}
}

@inproceedings{diskann,
 author = {Jayaram Subramanya, Suhas and Devvrit, Fnu and Simhadri, Harsha Vardhan and Krishnawamy, Ravishankar and Kadekodi, Rohan},
 booktitle = {Advances in Neural Information Processing Systems},
 editor = {H. Wallach and H. Larochelle and A. Beygelzimer and F. d\textquotesingle Alch\'{e}-Buc and E. Fox and R. Garnett},
 pages = {},
 publisher = {Curran Associates, Inc.},
 title = {DiskANN: Fast Accurate Billion-point Nearest Neighbor Search on a Single Node},
 url = {https://proceedings.neurips.cc/paper_files/paper/2019/file/09853c7fb1d3f8ee67a61b6bf4a7f8e6-Paper.pdf},
 volume = {32},
 year = {2019}
}

@inproceedings{baranchuk2018revisiting,
  title={Revisiting the inverted indices for billion-scale approximate nearest neighbors},
  author={Baranchuk, Dmitry and Babenko, Artem and Malkov, Yury},
  booktitle={Proceedings of the European Conference on Computer Vision (ECCV)},
  pages={202--216},
  year={2018}
}

\end{document}